\documentclass[aps,prx,twocolumn,superscriptaddress,showpacs,final,floatfix,longbibliography]{revtex4-2}
\usepackage[utf8]{inputenc}
\usepackage{hyperref}
\usepackage{float}
\makeatletter
\let\newfloat\newfloat@ltx
\makeatother
\usepackage{algorithm}
\usepackage{algpseudocode}
\makeatletter
\renewcommand{\fnum@algorithm}{\ALG@name~\thealgorithm}
\makeatother
\usepackage[toc,page]{appendix}
\usepackage{amsfonts}
\usepackage{amsmath}
\usepackage{amssymb}
\usepackage{amsthm}
\usepackage{mathtools}
\usepackage{physics}
\usepackage[dvips]{graphicx}
\usepackage{epstopdf}
\usepackage{booktabs}
\usepackage{tabularx}
\usepackage{multirow}
\usepackage{comment}
\usepackage{txfonts}
\usepackage{dsfont}
\usepackage{bbold}
\usepackage{subcaption}
\usepackage{mathrsfs}

\hypersetup{colorlinks=true, citecolor=blue, linkcolor=blue, urlcolor=blue}

\usepackage{xcolor}

\usepackage{cleveref}
\crefname{equation}{Eq.}{Eqs.}
\crefname{figure}{Fig.}{Figs.}
\crefname{observation}{Obs.}{Obs.}
\crefname{corollary}{Corollary}{Corollaries}
\crefname{lemma}{Lemma}{Lemmata}
\crefname{proposition}{Proposition}{Propositions}

\allowdisplaybreaks[4]

\DeclareMathOperator{\Var}{Var}
\renewcommand{\tr}{\mathrm{tr}}

\newcommand{\id}{\mathbb{1}}                        
\renewcommand{\vec}[1]{\boldsymbol{#1}}             

\newcommand{\shiftobs}[2]{\delta_{#2} #1}                 

\newcommand{\estr}{\tilde{\boldsymbol{\mathsf{r}}}}
\newcommand{\estrcomp}[1]{\tilde{\mathsf{r}}_{#1}}

\renewcommand{\Re}[1]{\mathrm{Re}[#1]}
\renewcommand{\Im}[1]{\mathrm{Im}[#1]}

\newcommand{\av}[1]{\langle{#1}\rangle}             
\newcommand{\exrho}[1]{\langle{#1}\rangle_{\varrho}} 

\newcommand{\varho}[1]{{\rm Var}(#1)}       
\renewcommand{\var}[2]{{\rm Var}_{#2}(#1)}       
\newcommand{\Cov}[2]{\mathrm{Cov}_{#2}(#1)}         
\newcommand{\CovM}[2]{\Gamma_{#2}(#1)}              
\newcommand{\CovMrho}[1]{\Gamma_{\varrho}(#1)}      

\newcommand{\be}{\begin{equation}}
\newcommand{\ee}{\end{equation}}
\newcommand{\bea}{\begin{equation}\begin{aligned}}
\newcommand{\eea}{\end{aligned}\end{equation}}

\begin{document}

\title{The uncertainty geometry of finite-dimensional position and momentum}

\author{Dimpi Thakuria}
\email{dimpi.thakuria@tuwien.ac.at}
\affiliation{Vienna Center for Quantum Science and Technology, Atominstitut, TU Wien, Vienna 1020, Austria}

\author{Shuheng Liu}
\email{liushuheng@pku.edu.cn}
\affiliation{State key Laboratory of Artificial Microstructure and Mesoscopic Physics, School of Physics, Frontiers Science Center for Nano-optoelectronics, Peking University, Beijing 100871, China}
\affiliation{Vienna Center for Quantum Science and Technology, Atominstitut, TU Wien, Vienna 1020, Austria}

\author{Giuseppe Vitagliano}
\email{giuseppe.vitagliano@tuwien.ac.at}
\affiliation{Vienna Center for Quantum Science and Technology, Atominstitut, TU Wien, Vienna 1020, Austria}

\author{Konrad Szymański}
\email{k.sz@quantumstat.es}
\affiliation{Research Center for Quantum Information, Slovenská Akadémia Vied, Dúbravská cesta 9, 84511 Bratislava, Slovakia}
\affiliation{Vienna Center for Quantum Science and Technology, Atominstitut, TU Wien, Vienna 1020, Austria}

\begin{abstract}
Uncertainty relations are usually stated as bounds on selected combinations of variances, but the full covariance matrix contains substantially richer information about the geometry of quantum state space and about the operational capabilities of quantum systems. Here we characterize the covariance matrices attainable by a finite-dimensional canonical pair of observables related by the discrete Fourier transform, the natural analogue of position and momentum in a finite Hilbert space. We combine analytic arguments with convex-geometric and semidefinite-programming methods based on joint numerical ranges to describe the admissible region through unitary invariants, in particular the trace and determinant of the covariance matrix. This provides a systematic way to identify extremal states, generalizing the notion of minimum-uncertainty states, and to quantify how the discrete uncertainty geometry approaches its continuous counterpart with increasing dimension. We further show that the resulting covariance-matrix characterization has direct consequences for applications: it yields accuracy bounds for multiparameter estimation protocols and separability criteria for finite-dimensional bipartite systems, including discrete analogues of continuous-variable EPR-type witnesses. Our results establish a systematic and versatile platform for connecting uncertainty relations, convex quantum geometry, metrology, and entanglement detection in finite-dimensional systems.
\end{abstract}

\date{\today}
\maketitle

\section{Introduction}

Uncertainty relations are among the most distinctive structural features of quantum theory. In their most common formulations, they quantify the impossibility of preparing states with simultaneously sharp values for incompatible observables, and they are most naturally expressed in terms of second moments and covariance matrices~\cite{Robertson1929,RobertsonPR1934}. Over the years, this perspective has been complemented by many alternative formulations, including entropic uncertainty relations~\cite{Wehner_2010Rev,ColesetAlRevModPhys}, characteristic-function approaches~\cite{RudnickiTascaWalbornCharUNCPRA2016}, and refinements involving monotone metrics and quantum Fisher information~\cite{CaianielloGuz1988QuantumFisherMetric,Petz_2002,gibilisco2008quantum,Andai_2008,tothfrowis2022}.
Historically, uncertainty relations were already central to the discussion of completeness in quantum mechanics initiated by Einstein, Podolsky, and Rosen and by Bohr's reply~\cite{EPR1935,bohrEPR}. Their utility is not restricted to historical or foundational arguments, though~\cite{werner2019uncertainty}: in modern quantum information, covariance matrices and uncertainty relations provide powerful tools
that find applications ranging from quantum metrology~\cite{TothApellaniz2014} to entanglement detection and quantification~\cite{hofman03,WernerWolf2001,Hyllus06,guhnecova,gittsovich08,GittsovichQuant10,LiuVitagliano2022}. For example, states that minimize variance-based uncertainty relations are intimately related to coherent and squeezed states~\cite{Jackiw1968,VVDodonov_2002,Trifonov94,Trifonov_1997,KitagawaUeda1993,Wineland1994Squeezed}, which play a central role in quantum optics and in precision measurements beyond the standard quantum limit, and, in many-particle systems they are entangled~\cite{Sorensen2001Many-particle,SoerensenMoelmer2001,TothApellaniz2014,pezzerev18,FriisVitaglianoMalikHuberReview19}. More generally, the covariance matrix is closely tied to metrological sensitivity through its relation to the quantum Fisher information~\cite{Pezze2009Entanglement,Petz_2002,PETZ_2011,yu2013quantum,toth2013extremal,liu2020quantum}.

The properties of quantum mechanics captured by uncertainty relations have led to the development of a rich mathematical subject, not restricted to lower bounds and sum of variances only.
This includes geometric descriptions of uncertainty regions and of the extremal states that bound them~\cite{Kechrimparis_2016,Kechrimparis_2017}, as well as state-independent and many-observable generalizations~\cite{deGuiseetal2018,Dammeier_2015,SchwonnekDammeierWerner2017,Szymanski_2019}. 
A natural next step is to characterize the full set of achievable covariance matrices, and in particular its boundary, capturing the extremal variance behavior~\cite{leka2013some,petz2014characterization,leka2014noteextremaldecompositioncovariances}. 

The most paradigmatic uncertainty relations often involve canonical position/momentum pairs, for historical reasons but also for their maximal incompatibility~\cite{Busch2007155}. 
In fact, complementary approaches focused on the measurement uncertainty, capturing the properties of joint approximate measurements are also centered mostly on position and momentum being the canonical example~\cite{werner2004uncertainty,RevModPhys.86.1261,Busch14}. We focus on the covariance-matrix viewpoint, directly defined by states and observables, within the framework of \emph{preparation uncertainty}, connecting algebraic incompatibility, geometry of state space, and experimentally accessible second moments.

Specifically, we focus on the finite-dimensional analogue of a canonical pair, which also in some sense share the maximal incompatibility property. These operators are obtained from one another through the finite Fourier transform, a standard construction in discrete phase-space methods~\cite{AVourdas_2004,vourdasbook}, which in particular also implies that the eigenbases of the two operators are maximally unbiased~\cite{DURT_2010}. Finite Fourier-related observables have appeared in several forms in the literature on discrete phase spaces and discrete uncertainty relations~\cite{MassarSpindel08,deGuiseetal2018}, but a systematic covariance-matrix analysis of their jointly achievable second moments is, to our knowledge, still missing. 

Understanding this problem is interesting both conceptually and practically. Conceptually, finite-dimensional Fourier pairs provide a controlled setting in which one can compare discrete and continuous canonical structures, which is a long-standing and still active subject of intense investigation, including definition of finite phase space~\cite{vstovivcek1984quantum,aldrovandi1990structure,AVourdas_2004,zhang2004analytic,vourdas2006analytic,atakishiyev2008discrete,BangBergerPRA2009,cotfas2011finite,marchiolli2012theoretical,albert2017general,vourdasbook} as well as finite-dimensional analogue of the harmonic oscillator~\cite{barker2000discrete,cotfas2011finite,albert2017general}. Practically, they are relevant whenever continuous quadratures are accessed only through finite resolution or coarse-grained measurements, as happens in realistic detection schemes; in such scenarios, complementary coarse-grained observables and their uncertainty relations become operationally meaningful~\cite{Rudnicki_2012,Toscano_2018,TascaetaklPRL1028,PauletalPRA2018,TascaRudnickietal2018PRAent}.
The shift from continuous variables provides a richer geometry, too: finite-dimensional operators have bounded spectra, unlike their continous variable counterparts, leading to compactness of the achievable covariance matrices, and the existence of nontrivial \emph{upper bounds} to uncertainty, completely absent in the CV scenario.

Our main goal is to fully describe the admissible preparation uncertainty region of a finite-dimensional canonical pair, by characterizing the boundary of the set of allowed covariance matrices. In particular, projecting the region of admissible covariance matrices into natural invariants such as trace and determinant we identify states corresponding to the boundary of the uncertainty region, i.e., extremal uncertainty states. These provide also a generalized set of states that are useful for applications, especially in metrology. In this way, we obtain a finite-dimensional uncertainty geometry that parallels several well-known features of the continuous-variable case while also displaying genuinely discrete effects.

This parallel also extends to the multipartite case: The characterization of the allowed covariance matrices of a single particle allows us to find entanglement criteria which provide a finite-dimensional analogue of the covariance matrix criterion, which is of paramount importance in CV systems as it allows to fully characterize and quantify entanglement of multi-mode Gaussian states~\cite{AdessoIlluminati2007,weedbrook2012gaussian}.

The paper is organized as follows. First we introduce some background methods: In \cref{sec:CM-basics}, we review the basic properties of covariance matrices and their connection to the Robertson-Schrödinger uncertainty relation, while in \cref{sec:finite-QP}, we introduce the finite-dimensional analogues of position and momentum operators, constructed via the finite Fourier transform. 

Afterwards, we present our results: In \cref{sec:numerical}, we describe numerical tools for characterizing the set of achievable covariance matrices, including joint numerical range methods and sampling algorithms applied to the canonical pair. 
In \cref{sec:applications}, we discuss applications to metrology and entanglement. Focusing on the multiparameter estimation of finite-dimensional displacement operators we firstly derive accuracy bounds coming from the quantum Fisher information matrix in \cref{subsec:prep_to_meas_to_bounds} and then discuss the achievability of those bounds via a multiparameter method of moments in \cref{sec:applications_MOM}. In \cref{sec:entanglement_criteria} we then consider a two-particle scenario and obtain entanglement criteria as another application of our methods. 
We find that states that resemble CV Gaussian states are detected even more efficiently than with typical methods based on mutually unbiased bases, improving the tolerance to thermal noise.

Finally, in \cref{sec:conclusions} we conclude, summarizing our findings and discussing further applications of our methods as an outlook.

\subsection{Covariance matrices and uncertainty relations}
\label{sec:CM-basics}

Given a quantum state $\varrho$ and two observables $A$ and $B$, we define the covariance 
\be
\Cov{A,B}{\varrho} := \exrho{AB} - \exrho{A}\exrho{B},
\ee
and the $2 \times 2$ covariance matrix as
\be\label{eq:posCM}
\CovMrho{A,B} = \begin{pmatrix}
\varho{A} & \Cov{A,B}{\varrho} \\
\Cov{B,A}{\varrho} & \varho{B}
\end{pmatrix},
\ee
where $\varho{A} = \Cov{A,A}{\varrho} =  \exrho{A^2} - \exrho{A}^2$ is the variance~\footnote{We use the non-symmetric covariance $\Cov{A,B}{\varrho} = \exrho{AB} - \exrho{A}\exrho{B}$, which makes $\Gamma$ Hermitian but generally not real. In the CV and Gaussian state literature, the symmetric covariance $\frac{1}{2}\exrho{AB + BA} - \exrho{A}\exrho{B}$ is more common.}.

The covariance matrix is: (i) \emph{hermitean}, i.e., $\CovMrho{\vec{A}} = 
\Gamma_\varrho^\dagger(\vec A)$, (ii) \emph{positive semidefinite}, $\CovMrho{A,B} \geq 0$ and (iii) \emph{concave} under mixing:
\be\label{eq:concavity}
\CovM{\vec{A}}{p\varrho_1 + (1-p)\varrho_2} \geq p\, \CovM{\vec{A}}{\varrho_1} + (1-p)\, \CovM{\vec{A}}{\varrho_2}.
\ee

A matrix version of the Robertson-Schr\"odinger uncertainty relation can be then given as the positivity of the (hermitean, but non-symmetric) covariance matrix, i.e.,
\be
\Gamma_\varrho(\vec{A}) = \Gamma_\varrho^S(\vec{A}) + i \Omega_\varrho(\vec{A}) \geq 0 ,
\ee
where we introduced the symmetric $\Gamma_\varrho^S = \Re{\Gamma_\varrho}$ and skew-symmetric $\Omega_\varrho = \Im{\Gamma_\varrho}$ components, as it is most commonly done in literature.

Under linear transformation $\vec{A} \mapsto \vec{B} = \Lambda \vec{A}$ with real matrix $\Lambda$ the covariance matrix transforms as
\be
\CovMrho{\vec{B}} = \Lambda\, \CovMrho{\vec{A}}\, \Lambda^T
\ee
In particular, the covariance matrix $\CovMrho{\vec{A}}$ can be more intuitively characterized by its unitary invariants, e.g., the coefficients of the characteristic polynomial
\be
\det(\CovMrho{\vec{A}} - \lambda \id) = \sum_{r=0}^{M} C_r^{(M)} (-\lambda)^{M-r} ,
\ee
where $C_r^{(M)}$ is the sum of all principal minors of order $r$. In fact, the positivity of the determinant yields the ubiquitous Robertson-Schrödinger uncertainty relation~\cite{Robertson1929,RobertsonPR1934}:
\be\label{eq:RobSchUR}
\varho{A} \varho{B} \geq \tfrac{1}{4} \left( |\exrho{[A,B]}|^2 + \left| \exrho{\{A,B\}} - 2\exrho{A}\exrho{B} \right|^2 \right).
\ee
This extends to $M$ observables: given a vector $\vec{A} = (A_1, \ldots, A_M)$, the $M \times M$ covariance matrix $[\CovMrho{\vec{A}}]_{jk} = \exrho{A_j A_k} - \exrho{A_j}\exrho{A_k}$ is positive semidefinite, and its positivity implies a hierarchy of scalar uncertainty relations~\cite{Trifonov_1998,DODONOV1980150}.
The overall characterization of the covariance matrix becomes somewhat simpler where only two observables are considered, which is the case we will also mostly focus in the following. In that case, in particular, the only relevant invariants are $C_1= \tr \CovMrho{\vec{A}}$ and $C_2 = \det \CovMrho{\vec{A}}$. Furthermore, for two observables the concavity inequality \eqref{eq:concavity} can always be saturated by a suitable pure-state decomposition: for any $\varrho$, there exists a decomposition $\varrho = \sum_k p_k \ketbra{\psi_k}$ such that $\CovMrho{A,B} = \sum_k p_k \CovM{A,B}{\psi_k}$~\cite{leka2013some,leka2014noteextremaldecompositioncovariances}.

Typically in literature, states that saturate the uncertainty relation in \cref{eq:RobSchUR} are dubbed minimum uncertainty states~\footnote{Note that in the literature, many definitions of (variance-based) uncertainty relations and corresponding saturating states have been studied (see, e.g., \cite{Trifonov_2000}). Here, we are focusing on \cref{eq:RobSchUR} for the case of two observables.}. These essentially correspond to states for which $\det \CovMrho{\vec{A}} = 0$, and, for the case of two observables, have been studied extensively in general. See, e.g., Refs.~\cite{DODONOV1980150,Trifonov_1997}. More details on how to find them are given in \cref{sec:minunc}.

\subsection{Finite-dimensional canonical pair of observables}
\label{sec:finite-QP}

We consider finite-dimensional analogues of the canonical quadratures, which are defined via finite Fourier transform~\cite{AVourdas_2004}. In a Hilbert space of dimension $d$, we define the position operator in the computational basis as 
\be
Q := \sqrt{\frac{2\pi}d} \sum_{n=-(d-1)/2}^{(d-1)/2} n \ketbra{n}, \label{eq:qdef}
\ee
with eigenvalues centered at zero\footnote{The reason for the prefactor $\sqrt{2\pi/d}$ in \eqref{eq:qdef} is that with this choice, for large $d$ the low-lying spectrum of $\frac{1}{2}(Q^2 + P^2)$ approximates that of the CV harmonic oscillator, whose eigenvalues are $n + \frac{1}{2}$ for $n = 0, 1, 2, \ldots$ 
}. The momentum operator is obtained via the finite Fourier transform:
\be
\label{eq:pandfdef}
P := F Q F^\dagger, \qquad F := \frac{1}{\sqrt{d}} \sum_{j={-(d-1)/2}}^{(d-1)/2}\sum_{k=-(d-1)/2}^{(d-1)/2} \omega_d^{jk} \ketbra{j}{k},
\ee

where $\omega_d = e^{2\pi i/d}$ is the primitive $d$-th root of unity. By construction, $Q$ and $P$ are two {\it complementary observables} that have the same spectrum and mutually unbiased eigenbases~\cite{DURT_2010}:
\be
\ket{\tilde j} = \frac 1 {\sqrt d} \sum_{k,l} \omega_d^{kl} \ket{k} \braket{l}{j} = \frac 1 {\sqrt d} \sum_{k} \omega_d^{jk} \ket{k} ,
\ee
where we used the fact that the canonical basis is orthonormal, namely $\bra{l} \ket{j} = \delta_{lj}$.
Thus, the overlap between an eigenvector $\ket{k}$ of $Q$ and $\ket{\tilde j}$ of $P$ is
\be
\braket{k}{\tilde j} = \frac 1 {\sqrt d} \omega_d^{jk} ,
\ee
which indeed shows that the two bases are mutually unbiased.

The matrix elements of $P$ in the $Q$ eigenbasis can be written as analogous of ``derivatives'', namely making use of the following functions~\cite{AVourdas_2004}
\be
\Delta_m(x) := \partial_x^m \Delta_0(x) = \partial_x^m \left( \frac 1 d \sum_n \omega^{nx} \right) = \frac 1 d \sum_n \left( \frac{2\pi i}{d} n \right)^m \omega^{nx}_d . 
\ee
With this definition we obtain
\be
P = \sqrt{\frac{2\pi}d} \sum_{nm} \sum_{k} k \frac{\omega_d^{k(n-m)}} d \ketbra{n}{m} = \frac 1 d \left( \frac{d}{2\pi i} \right) \Delta_1(n-m)  ,
\ee
and more explicitly
\begin{align}
\label{comm_1}
    Q_{jk}&=\sqrt{\frac{2\pi}d} j \delta_{jk} \\ 
    P_{kl}&=\bra{k}FQF^\dag \ket{l}=\left\{
    \begin{array}{cc}
       0  & \text{for}\quad k=l\nonumber \\
      -\sqrt{\frac{2\pi}d}\frac{i}{2}\frac{(-1)^{(k-l)}}{\sin(\frac{\pi}{d}(k-l))} & \text{for}\quad k\neq l
    \end{array}
    \right.
\end{align}
Therefore,  the commutator becomes
\be
[Q,P]_{kl}=\left\{
    \begin{array}{cc}
       0  & \text{for}\quad k=l\nonumber \\
      - i \frac{\pi}{d} (-1)^{(k-l)} \frac{(k-l)}{\sin(\frac{\pi}{d}(k-l))}  & \text{for}\quad k\neq l
    \end{array}
    \right.
\ee 
To complete the picture, it is worth pointing out that it is actually more common to introduce finite phase space by
defining the discrete analogous of displacement operators as follows 
\be
\begin{aligned}
X &= e^{\sqrt{\frac{2\pi}{d}} iP} = \sum_{k} \omega^{k}_d \ketbra{k}{k} , \\
Z &= e^{-\sqrt{\frac{2 \pi  }{d}}iQ} = \sum_{k} \omega^{-k}_d \ketbra{\tilde k}{\tilde k} .
\end{aligned}
\ee
The action of the displacement operators in the position and momentum eigenbases are given by a cyclic permutation of the elements respectively~\cite{AVourdas_2004,DURT_2010,vourdasbook}.

Note also that these operators 
have mutually unbiased eigenbases by construction. From them one can define Heisenberg-Weyl operators as follows 
\be\label{eq:DiscreteDisplacements}
D(l,n,m) = \omega^l_d X^n Z^m ,
\ee
where $(l,n,m)$ are three integer numbers that are smaller than $d-1$, i.e., $0\leq l,n,m \leq d-1$. 

Note that for $d=2$ the Heisenberg-Weyl group is formed (essentially) by the three Pauli matrices plus the identity matrix.
Besides this, it is also possible to finite-dimensional version of frames and use it to quantize states and operators~\cite{zhang2004analytic,cotfas2011finite}.
It is worth pointing out that this construction of finite frames is based on looking for the analogous of the CV coherent states, which in CV are also minimum uncertainty states.
In finite dimensional systems, these analogous of coherent states need not strictly be minimum uncertainty states, but it has been shown that they are at least close to have such a property~\cite{cotfas2011finite}.

\section{Characterization of covariance matrix for finite canonical pair}
\label{sec:numerical}
\begin{figure*}
  \centering
  \begin{subfigure}[b]{0.34\linewidth}
    \includegraphics[width=\linewidth]{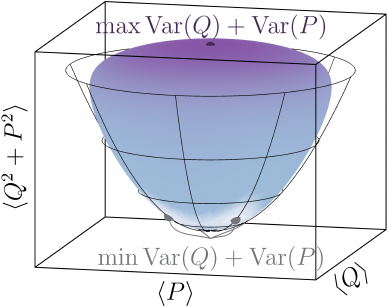}
    \caption{$d=4$}
  \end{subfigure}\hspace{10mm}%
  \begin{subfigure}[b]{0.33\linewidth}
    \includegraphics[width=\linewidth]{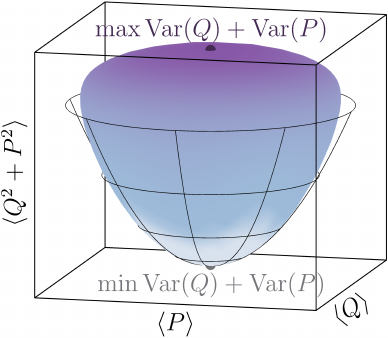}
    \caption{$d=5$}
  \end{subfigure}
  
    \caption{Joint numerical ranges (shaded volumes) of $Q$, $P$, and $T=Q^2+P^2$ for dimensions $d=4$ and $d=5$. Wireframes show level sets of the sum of variances $\var{Q}{}+\var{P}{}$, which are paraboloids of revolution in this space. Dots mark the minimum and maximum variance states: for odd $d$, the minima lie on the $\langle Q\rangle=\langle P\rangle=0$ axis, while for even $d$ they form a fourfold-symmetric orbit away from it (two of four are visible in (a)). The maximum always lies on the central axis.}
    \label{fig:JNR-QPT}
\end{figure*}

In this section we present the tools used to analyze the covariance matrix region of a pair of observables. While the methodology applies to arbitrary pairs, we restrict attention  to the finite-dimensional canonical pair $(Q,P)$ introduced in \cref{sec:finite-QP}: a \emph{maximally incompatible pair}, with mutually unbiased eigenbases. Analogous characterizations have been carried out for $d=2$ and partially for $d=3$~\cite{busch2019arxiv}.

\subsection{Geometric considerations and convex geometry}
To characterize the full region of allowed covariance matrices of a canonical pair, we consider a geometric method based on the joint numerical range.
The idea is to probe the set of all achievable expectation values for a suitable tuple of operators, from which the covariance matrix can be reconstructed. For the two observables $Q$ and $P$, define the Hermitian operators
\begin{equation}
\label{eq:TXYZ-ops}
\begin{aligned}
T &:= Q^2 + P^2, & G_3 &:= Q^2 - P^2, \\
G_1 &:= \{Q,P\}, & G_2 &:= -i[Q,P].
\end{aligned}
\end{equation}
These, together with $Q$ and $P$ themselves, form a six-tuple whose joint numerical range encodes all information needed to reconstruct covariance matrices:
\begin{equation}
\mathcal{J} := \big\{ (\av{Q}, \av{P}, \av{T}, \av{G_1}, \av{G_2}, \av{G_3}) : \varrho \geq 0,\, \tr\varrho = 1 \big\} \subset \mathbb{R}^6.
\end{equation}
This is because a general covariance matrix can be written as
\begin{equation}
    \Gamma = \frac12 \begin{pmatrix}t+g_3 & g_1+i g_2\\g_1-i g_2 & t-g_3\end{pmatrix} - \begin{pmatrix}q^2 & qp\\qp & p^2\end{pmatrix},
\end{equation}
with $(q,p,t,g_1,g_2,g_3)\in\mathcal{J}$. As an affine image of the convex compact set of density matrices, $\mathcal{J}$ is a convex compact body, equal to the convex hull of its pure-state image. By the Carath\'eodory's theorem \footnote{Carath\'eodory's theorem: for any subset $P \subset \mathbb{R}^k$, every point in $\mathrm{conv}(P)$ can be written as a convex combination of at most $k+1$ points of $P$.}, every point in $\mathcal{J}$ is a convex combination of at most seven pure-state tuples, and is therefore realized by some state of rank at most seven. In fact, a much stronger statement holds for such numerical ranges: rank-$2$ density matrices already suffice to realize every point inside $\mathcal{J}$ \cite{au1979remark}.  Consequently, the entire set of covariance matrices is attained by states of rank at most $2$, regardless of the Hilbert space dimension and details of the observables.

A projection of this set into 3 dimensions provides information about the extremal values of the trace of $\Gamma$, equal to the sum of variances $\var{Q}{}+\var{P}{}$. This can be done with the following joint numerical range 
\begin{equation}
    \mathcal{J}_3 = \big\{ (\av{Q}, \av{P}, \av{T}) : \varrho \geq 0,\, \tr\varrho = 1 \big\} \subset \mathbb{R}^3.
\end{equation}
In a fashion similar to previous works \cite{Szymanski_2019,schwonnek2017state}, the minimum sum of variances can be determined using methods of polynomial equation solving a numerical optimization over the boundary of $\mathcal{J}_3$; details of the polynomial approach can be found in \cite{Szymanski_2019}. The calculations rely on the variational formulation
\begin{align}
\tau_{\min}
&:=\min_\psi \bigl(\Var_\psi(Q)+\Var_\psi(P)\bigr) \nonumber\\
&=
\min_{q,p\in\mathbb R}\ \min_\psi
\bra{\psi}\Bigl[(Q-q\mathbb 1)^2+(P-p\mathbb 1)^2\Bigr]\ket{\psi}.
\label{eq:U-def-paper}
\end{align}
The inner minimization is an eigenvalue problem, and
\begin{equation}
\tau_{\min}=\min_{q,p}\lambda_{\min} \bigl((Q-q \mathbb 1)^2+(P-p\mathbb 1)^2\bigr).
\label{eq:U-groundstate-paper}
\end{equation}
Then, the outer minimization can be performed analytically, as it reduces to the vanishing of the derivatives of the characteristic polynomial of $(Q-q \mathbb 1)^2+(P-p\mathbb 1)^2$.
This has to be done separately for every dimension; empirically, the optimal points minimizing the sum of variances for odd Hilbert space dimension ($d=3,5,\ldots$) have $\av{Q}=\av{P}=0$, so that for these cases the minimum sum of variances reduces to the minimum eigenvalue of $T$.

For even dimensions, on the contrary, the minimal sum of variance points do not lie on the $\av{Q}=\av{P}=0$ axis. There is still some structure to them, due to the fourfold rotational symmetry of $\mathcal{J}_3$ stemming from the symmetries of the operators defining it: under conjugation by the Fourier transform \eqref{eq:pandfdef}, the observables exchange:
\begin{equation}
    \begin{aligned}
    F Q F^\dagger = P, \quad F P F^\dagger = -Q, \quad
    F T F^\dagger = T,
    \end{aligned}
\end{equation}
which means that the joint numerical range is symmetric with respect to rotations by $\pi/2$ around the vertical $\langle T \rangle$ axis. Thus, every point realizing the minimal sum of variances not lying on this axis is a part of an orbit of four solutions.

The maximum sum of variances lies at the topmost point of $\mathcal{J}_3$ on the $\av{Q}=\av{P}=0$ axis. Indeed, the objective $\av{T}-\av{Q}^2-\av{P}^2$ is strictly concave in $(\av{Q},\av{P})$, and fourfold symmetry of $\mathcal{J}_3$ combined with strict concavity forces the maximizer onto the symmetry axis -- see Fig. \ref{fig:JNR-QPT}. The maximal sum of variances therefore equals the maximal eigenvalue of $T$, regardless of the Hilbert space dimension. 

Both minimal and maximal sums of variances are realized by pure states: for the joint numerical range of three Hermitian operators on a Hilbert space of dimension $d \geq 3$, the convex hull of pure-state expectation tuples already coincides with the full mixed-state JNR \cite{au1979remark}, and the $d=2$ case is convex by the particular structure of the problem.

Note also that the eigenstates of $T$ are related to finite-dimensional Harmonic oscillator eigenstates, which have also attracted considerable attention broadly over time in various contexts~\cite{mehta1987eigenvalues,barker2000discrete,vourdas2006analytic,atakishiyev2008discrete,Cotfas_2012,vourdasbook}.

\begin{figure}
    \centering
     \includegraphics[width=.6\linewidth]{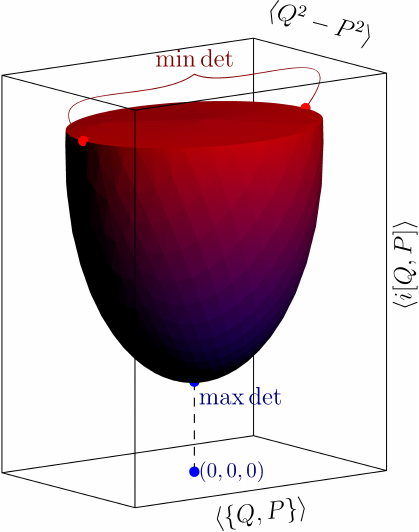}

    \caption{Joint numerical range of $i[Q,P],\{Q,P\},Q^2-P^2$ for states restricted to fixed values of $\av{Q^2+P^2}, \av{Q},\av{P}$. For each such choice of constraints, points on the boundary farthest from the origin correspond to minimal determinant, while points closest to the origin correspond to maximal determinant.}
    \label{fig:JNR-crosssections}
\end{figure}

As the covariance matrix is necessarily positive semidefinite, from the eigenvalue bounds $\lambda_1,\lambda_2\ge 0$ we get $0 \le \det \Gamma \le \frac14 (\tr \Gamma)^2$. Further constraints can be provided by another numerical range. To visualize the resulting set we can fix three variables, for example 
$t:=\av{T}, \av{Q}=\av{P}=0$~\footnote{The expectation $\av{Q},\av{P}$ can be set to arbitrary values within the projection of $\mathcal{J}$ onto the $(\av{Q},\av{P})$ plane; the reasoning following does not change substantially, here the zero is chosen for simplicity.}. Then, the expectation values of the remaining three ($G_1, G_2, G_3$ defined in \eqref{eq:TXYZ-ops}) define a region in $\mathbb{R}^3$:
\begin{equation} \label{eq:Jt}\begin{aligned}
\mathcal J_t
:=
\bigl\{
(\av{G_1},\av{G_2},\av{G_3}):
\ \av{T}_\varrho=t,\ \av{Q}_\varrho=\av{P}_\varrho=0,\\
\ 
\varrho\ge0,\ \tr\varrho=1
\bigr\},\end{aligned}
\end{equation}
which can be efficiently approximated by semidefinite optimization: a point on the boundary with normal $(n_1, n_2, n_3)$ corresponds to the optimum of
\begin{equation}
 \tr \bigl[\varrho(n_1 G_1+n_2 G_2+n_3 G_3)\bigr],
 \end{equation}
 with the constraints equivalent to those appearing in Eq. \eqref{eq:Jt}.

At fixed trace, the determinant depends only on the radius in $(g_1,g_2,g_3)$-space:
\begin{equation}
\det\Gamma_\varrho(Q,P)=\frac14\bigl(t^2-r^2\bigr),
\qquad
r^2:=g_1^2+g_2^2+g_3^2.
\label{eq:det-from-radius-paper}
\end{equation}
Note that positivity of $\Gamma$ already forces $r \leq t$ within $\mathcal{J}_t$. Hence, for fixed $t$, minimal determinant corresponds to points in $\mathcal J_t$ farthest from the origin, while maximal determinant corresponds to points closest to the origin (see Fig. \ref{fig:JNR-crosssections}). If the origin belongs to $\mathcal J_t$, then the upper bound
\begin{equation}
\delta_{\max}(t)=\frac{t^2}{4}
\end{equation}
is attained, corresponding to an isotropic covariance matrix.

\subsection{Trace-determinant regions via direct optimization}
The convex-geometry methods shown above could be used to provide the full description of the trace-determinant region, but due to their computational complexity they are better suited for theoretical understanding of its structure; other numerical methods are needed for systematic sampling.

Therefore, for the analysis below, we use a numerical algorithm that
directly optimizes (minimizes or maximizes) the determinant subject to a
trace constraint, with the admissible trace range taken from the bounds
derived in the previous section.

\begin{figure}
    \begin{center}
    \includegraphics[width=1.0\linewidth]{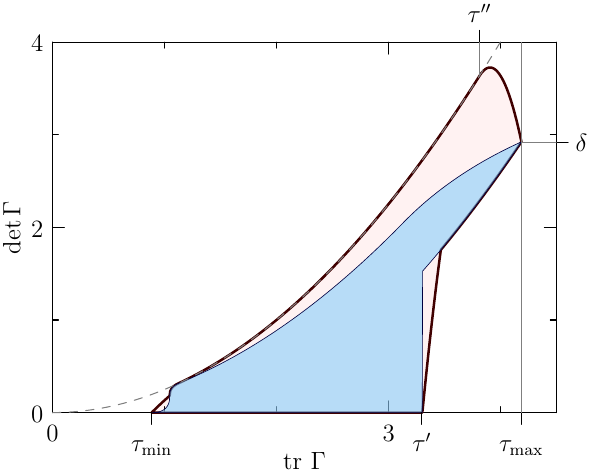}
    \caption{Allowed region in the $(\tr\Gamma_\varrho,\det\Gamma_\varrho)$-plane  for the finite-dimensional canonical pair $(Q,P)$ for dimension $d=3$. Two cases are depicted: mixed quantum states (outer region with thick outline and light red background) and pure states (light blue). The pure state region is strictly contained in the mixed-state region.}
    \label{fig:trdetregionfinite}
    \end{center}
\end{figure} 

The goal is to characterize the trace-determinant region shown in
\cref{fig:trdetregionfinite}.
The algorithm relies on a standard parameterization of rank-$k$ states
in dimension $d$ through an arbitrary complex matrix $A$ of size
$d\times k$:
\begin{equation}
    \varrho := \frac{A A^\dagger}{\tr (A A^\dagger)}.
\end{equation}
This form automatically enforces positivity and unit trace, and the rank
constraint is imposed by the choice of $k$.
In practice we focus on $k=1$ (pure states) and $k=d$ (full-rank states). 

\begin{algorithm}[h]
\caption{Trace-constrained determinant extremization}
\label{alg:trace-det}
\begin{algorithmic}[1]
\Require dimension $d$; rank $r$; target trace $t$; direction $\in\{\min,\max\}$;
number of restarts $N$
\Ensure extremal $\det\Gamma$ at trace $t$ over rank-$r$ states
\State Parametrize $\varrho = AA^\dagger / \tr(AA^\dagger)$ with $A \in \mathbb{C}^{d \times r}$
\For{$N$ random initializations of $A$}
  \State Locally optimize (minimize or maximize, depending on the direction argument) $\det\Gamma_\varrho(Q,P)$,        subject to $\tr\Gamma_\varrho(Q,P) = t$
\EndFor
\State \Return best (minimal/maximal) value across restarts.
\end{algorithmic}
\end{algorithm}

This optimization is non-convex and we use multiple random restarts to mitigate local minima.
A typical example of numerically obtained region in the $(\tr \Gamma,\det\Gamma)$-plane is shown in \cref{fig:trdetregionfinite}. In particular, the pure-state region is strictly contained in the mixed-state region: there exist pairs $(\tr\Gamma,\det\Gamma)$ that are realized by mixed states only. We also observe that, except for $d=3$, the state minimizing the sum of variances does not have zero determinant: the minimum-determinant boundary of the pure-state region lifts slightly from zero at the smallest accessible trace. The lift is small \footnote{For $d=5$, it is equal to $-\frac{\left(-146925+64903 \sqrt{5}+5 \sqrt{6442 \left(210475-94119 \sqrt{5}\right)}\right) \pi ^2}{322100}\approx 2.16\times 10^{-3}$, as it can be calculated using the eigenstate to minimum eigenvalue of $T$.} and decays rapidly with dimension. 

The $d=3$ case is exceptional, and partial characterization of the $(\tr \Gamma,\det\Gamma)$-plane can be given analytically, as we show below.

\subsection{Extremal states in $d=3$}

For $d=3$, the minimal sum of variances is $\tr \Gamma_{\rm min} = \frac{2}{9} \left(3-\sqrt{3}\right) \pi$ and is attained by a state having zero determinant, i.e., a minimum uncertainty state (cf. \cref{sec:minunc}). In this case, i.e., for $d = 3$, the minimum uncertainty states can be constructed explicitly. 
First, an eigenstate of $Q + iP$ (analogous to the ``vacuum state'' in CV) is given by
\be
\ket{0_A} = \frac{1}{\sqrt{6 + 2\sqrt{3}}} \begin{pmatrix} 1 \\ 1 + \sqrt{3} \\ 1 \end{pmatrix}.
\ee
This state achieves the minimal trace and has equal variances of $Q$ and $P$.

Applying a squeezing-like unitary 
\be
U(r):=e^{-i\xi K} \quad \text{with} \quad K = \frac{1}{2}(QP + PQ)
\ee
generates a family of minimum uncertainty states $\ket{\psi(\xi)}$ parametrized by $\xi \in \mathbb{R}$. 
Explicitly this family of states is given by
$$
|\psi(\xi)\rangle = e^{-i\xi K}|0_A\rangle = \frac{1}{\sqrt{6+2\sqrt{3}}} \begin{pmatrix} x(\alpha) \\ y(\alpha) \\ x(\alpha) \end{pmatrix}, 
$$
where $\alpha = \frac{\sqrt{2}\pi \xi}{3\sqrt{3}}$ and 
\bea
x(\alpha) &= \cos\alpha + \frac{\sqrt{2}(1+\sqrt{3})}{2}\sin\alpha, \\ 
y(\alpha) &= (1+\sqrt{3})\cos\alpha - \sqrt{2}\sin\alpha.
\eea
These are analogue of single mode squeezed states in CV.

The trace along this family is
\be
\tr \Gamma(\xi) = \frac{2}{3} \pi  \left(1-\frac{1}{\sqrt{3}} \cos \left(\sqrt{\frac{8}{27}} \pi  \xi \right)\right),
\ee
with $\det \Gamma(\xi) = 0$ throughout. See \cref{app:d3exampledetails} for further details.
These states fully characterize the bottom line in the plot in \cref{fig:trdetregionfinite}. However, for larger dimension this simple intuitive picture of squeezed states corresponding to minimum uncertainty states does not hold anymore.

\section{Applications} \label{sec:applications}

\subsection{From preparation uncertainties to accuracy bounds in (multi)parameter-estimation}
\label{subsec:prep_to_meas_to_bounds}

As a first application, in this subsection we connect the \emph{preparation uncertainty} of a probe state, quantified by covariances of the generators of a (unitary) imprinting dynamics, to \emph{accuracy bounds} in (multi-)parameter estimation.
It is in fact well known that uncertainty relations acquire a direct operational meaning in quantum metrology, that translates into fundamental limits on attainable estimation accuracy once the parameter is encoded through the corresponding generator(s). In this sense, search for saturating states is not an abstract exercise: it identifies the probe preparations that allow achieving (or closely approaching) the quantum Cram\'er--Rao bound, which is the ultimate precision bound.
 
In the single-parameter, pure-state, unitary setting
$\ket{\psi_\theta}=e^{i\theta H}\ket{\psi_0}$,
the quantum Cram\'er--Rao bound (QCRB) takes the uncertainty-relation form
\be \label{eq:qcrb}
(\Delta \Theta)^2\ge \frac{1} {\nu F^Q_{\psi_\theta}(H)}=\frac1 {4\nu  \Var_{\psi_\theta}(H)} ,
\ee
where $\Theta$ is a generic {\it unbiased estimator} of $\theta$, $\nu$ is the number of i.i.d. repetitions of the experiment and $F^Q_{\psi_\theta}(H)$ is the {\it Quantum Fisher Information}, which for pure states and unitary encodings equals four times the variance of the generator on the state. 
Typically, being the QCRB valid in a local (point) estimation framework, one restricts to a neighborhood of the point $\theta= 0$ and concludes that a preparation with a large $\Var_{\psi_0}(H)$ is the \emph{resource} that sets the best possible sensitivity, which means that the best probe state is given by a superposition between the maximal and minimal eigenstates of $H$, i.e., ${\rm argmax}_{\ket{\psi_0}} \Var_{\psi_0}(H) = \tfrac 1 {\sqrt 2} (\ket{h_{\max}} + \ket{h_{\min}}) := \ket{\psi_{\rm opt}}$. 

Let us now consider the multi-parameter scenario and observe how to connect the preparation uncertainty of several variables into bounds on the accuracy of estimation of multiple parameters at once. For concreteness, let us focus once again on a finite-dimensional canonical pair $\vec R= (Q,P)$ as our phase space variables and
thereby consider a two-parameter ``displacement'' encoding
\be
\ket{\psi_{\vec r}} = e^{i (r_1 Q + r_2 P)} \ket{\psi_{0}} ,
\ee
with $\vec r = (r_1, r_2)$ being our vector of (real) parameters, and for simplicity let us assume we want to maximize the accuracy of the estimation around $\vec r = (0,0)$.
Thus, here the canonical pair plays the role of the generators of the (unitary) encoding dynamics.

For any (locally) unbiased (vector of) estimator(s) $\estr$ obtained from $\nu$ independent repetitions, the quantum Cram\'er--Rao bound yields the matrix inequality
\be
\mathrm{Cov}(\estr)
\;\succeq\;
(\nu F^Q_{\psi_{\vec r}}(\vec R))^{-1} = \left(4 \nu \,  \Gamma^S_{\psi_{\vec r}}(\vec R)\right)^{-1} ,
\label{eq:qcrb_matrix}
\ee
where $\mathrm{Cov}(\estr)$ is the covariance of the estimators and $F^Q_{\psi_{\vec r}}(\vec R)$ is the quantum Fisher information matrix (QFIM) associated to the unitary encoding generated by $\vec R$. In the present case of unitarily encoded pure states the QFIM equals four times the symmetric covariance matrix $\Gamma^S_{\psi_{\vec r}}(\vec R)=\mathrm{Re}[\Gamma_{\psi_{\vec r}}(\vec R)]$ of the generators~\cite{Helstrom1976,BraunsteinCaves1994}\footnote{Note that in the multiparameter case a more general bound called Holevo-Cram\'er-Rao was established~\cite{holevo2011probabilistic}.}. 
In the multiparameter case this bound in not always saturable. The condition for saturability in the pure state with unitary encoding becomes
\begin{equation}
\label{saturability_cond}
\big[J_{\psi_{\vec r}}(\vec H)\big]_{\mu\nu}
= -2i\bra{\psi_{\vec r}}[H_\mu,H_\nu]\ket{\psi_{\vec r}} = 0 ,
\end{equation}
with the $H_k:=
i\,U^\dagger\,\partial_k U$ being the generators of the dynamics. 
See also \cref{app:multi_metrology}.

In our case, considering small displacements $\vec r \simeq (0,0)$, we have $\vec H= \vec R$. Note that the saturability condition in this case reduces to the imaginary part of the covariance matrix being zero, i.e., $J_{\varrho}(\vec H)=\Omega_\varrho (\vec R)=0$
in this case.
The bound in \cref{eq:qcrb_matrix} is a matrix inequality, that can be therefore practically evaluated by tracing it with some positive semidefinite weight matrix $W$, which turns it into a scalar condition
\be\label{eq:QCRboundscalar}
\tr\left(W \mathrm{Cov}(\estr) \right) \geq \frac 1 {4\nu} \tr \left( W (\Gamma^S_{\psi_{0}}(\vec R) )^{-1}\right) .
\ee
Note that the matrix inequality in \eqref{eq:qcrb_matrix} holds if and only if all such scalar conditions with positive definite $W$ hold.
In particular, considering the common choice $W=\id$, which treats on equal footing all parameters leads to 
\be\label{eq:QCRboundfortrace}
\tr\left(\mathrm{Cov}(\estr) \right) \geq  \frac 1 {4\nu} \tr \left(  (\Gamma^S_{\psi_{0}}(\vec R) )^{-1}\right)  ,
\ee

Thus, in order to minimize the trace of the covariance matrix of the estimators, a useful probe state is potentially given by the one that minimizes the trace of the inverse covariance matrix.  
In case of the $2\times 2$ matrix $\Gamma^S_{\psi_{0}}(\vec R)$ that we are discussing, the trace of its inverse, $(\Gamma^S_{\psi_{0}}(\vec R) )^{-1}$
can be expressed as $\tr(\Gamma^S_{\psi_{0}}(\vec R) ) / \det (\Gamma^S_{\psi_{0}}(\vec R) )$. Therefore, the optimal (pure) probe state is the one that minimizes this ratio, leading to
\be
\label{A_d}
\mathcal A_d =\frac{1}{4} \min_{\psi \in \mathbb{C}^d} \tr \left(  (\Gamma^S_{\psi}(\vec R) )^{-1}\right) ,
\ee
which represents the maximal accuracy potentially achievable by a single $d$-dimensional system for this task. 
On top of that, one can also impose the saturability condition in \cref{saturability_cond} as an additional constraint in the optimization, thereby obtaining:
\be
\begin{aligned}
    \label{A_dc}
\mathcal A_d^c &=\frac{1}{4} \min_{\psi \in \mathbb{C}^d} \tr \left(  (\Gamma^S_{\psi}(\vec R) )^{-1}\right) \\
\text{such that} &\quad  J_{\psi}(\vec R) = 0  ,
\end{aligned}
\ee

 \begin{figure}
    \centering
   \includegraphics[width=1.25\linewidth,trim=3cm 9cm 0.7cm 9.6cm,
clip]{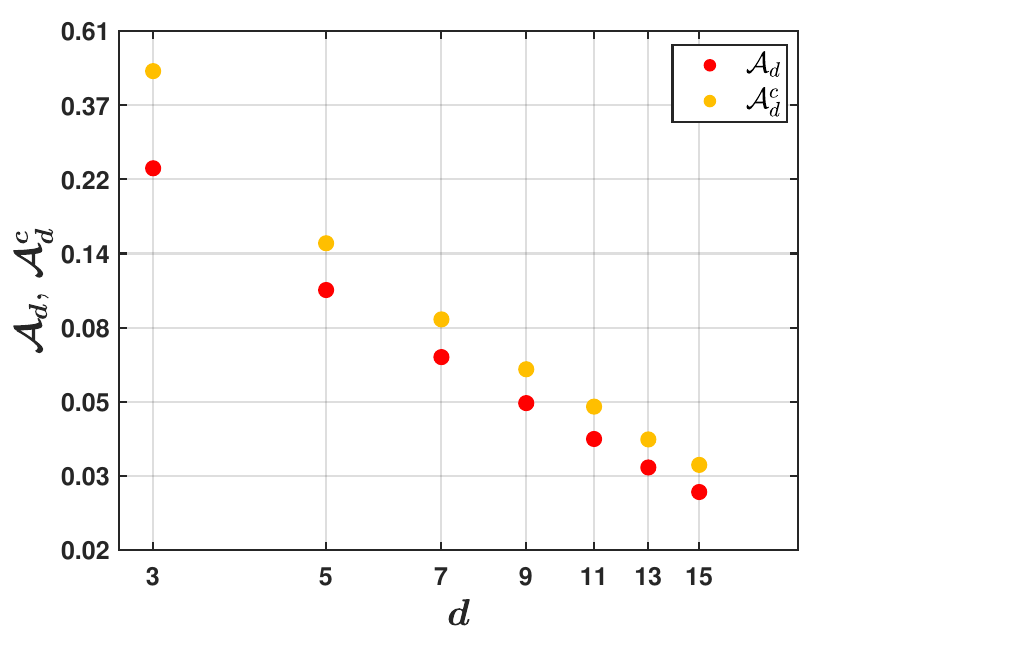}
    \caption{Value of $\mathcal A_d$, $\mathcal A^c_d$ as  functions of dimension in a log-log scale. The quantity  $\mathcal A_d$, represented by the red dots, provides a lower bound
    to the trace of the covariance matrix of the two estimators $\estr = (\estrcomp{1}, \estrcomp{2})$ (cf. \cref{eq:QCRboundfortrace}), and corresponds to optimization performed without imposing the saturability constraint. In contrast, the quantity $\mathcal A^c_d$ represented by the orange dots, corresponds to optimization with the saturability constraint imposed.}
    \label{fig:fig-accuracy vs dim 2 parameter}
\end{figure}

In \cref{fig:fig-accuracy vs dim 2 parameter} we plot both the values of $\mathcal A_d$ and $\mathcal A_d^c$ from \cref{A_d,A_dc} respectively as functions of the dimension $d$ in a log-log scale. The states that achieve these bounds  
can be readily determined using our method of characterizing covariance matrices in terms of their trace and determinant. In particular, the state $\ket{\psi_0}$ achieving the overall bound $\mathcal A_d$ lies on the upper boundary defined by the maximal determinant of $\Gamma^S_{\psi_{0}}(\vec R)$ for a fixed trace, which corresponds to pure states (cf. \cref{fig:trdetregionfinite}). Indeed, for a given trace value, maximizing the determinant minimizes the trace-to-determinant ratio, thereby optimizing the R.H.S of \cref{eq:QCRboundfortrace} at some trace value. 
Notably, in this multi-parameter setting, these optimal states are not as intuitively clear as in the single-parameter case, and moreover they do not automatically satisfy the saturability condition
\eqref{saturability_cond}, which means that indeed the best QCRB that is also saturable is a bit higher than the overall minimum, as it can be seen in \cref{fig:fig-accuracy vs dim 2 parameter}.

From our analysis, we observe, as expected that the minimimal QCRB increases with the dimension, and by fitting the log-log plot with a linear function we observe a scaling which is roughly in between a shot-noise scaling with $1/d$ and a Heisenberg scaling $1/d^2$. Note that in that case, it is also known how to find actual measurements that would allow to achieve the limiting accuracy, at least asymptotically using the maximum likelihood estimator~\cite{Matsumoto_2002,pezze_etal_PRL_multiphases}. 
Note that similarly, for other weight matrices one can define different scalar figures of merit, that might contain some asymmetry in the two parameters, and analyze what is the best achievable bound, depending on the given weighted trace of the inverse covariance matrix.

\subsection{Preparation and measurement uncertainties versus estimation accuracy}
\label{sec:applications_MOM}

Aside from the probe state, another ingredient that plays a central role is the {\it measurement}. Let us discuss this once again first in the single-parameter scenario.
A natural way to ask in a practical scenario how the Cram\'er-Rao bound can be saturated by an explicit measurement is to consider estimating $\theta$ from the mean of an observable $M$, quantifying the accuracy via error propagation~\cite{TothApellaniz2014,pezzerev18}: 
\begin{equation}
(\Delta \Theta)^2
=\frac 1 \nu \frac{\Var_{\psi_0}(M)}{\left|\partial_\theta\langle M\rangle_\theta\right|_{\theta=0}^2}
=\frac 1 \nu \frac{\Var_{\psi_0}(M)}{\bigl|\langle[M,H]\rangle_{\psi_0}\bigr|^2}\,,
\label{eq:errprop_intro}
\end{equation}
where $\partial_\theta\langle M\rangle_\theta|_{\theta=0}=i\langle[H,M]\rangle_{\psi_0}$. 

Comparing \eqref{eq:errprop_intro} with the QFI expression \eqref{eq:qcrb} shows that saturating the QCRB through a (single-shot) measurement of $M$ requires the state to saturate the (Schr\"odinger--)Robertson uncertainty relation for the pair $(M,H)$, i.e.
\begin{equation}
\Var_{\psi_0}(H)\,\Var_{\psi_0}(M)
=\tfrac14\bigl|\langle[M,H]\rangle_{\psi_0}\bigr|^2 .
\label{eq:MU_condition_intro}
\end{equation}
Thus, the states that realize equality in this scheme are precisely minimum-uncertainty states between the generator of the dynamics $H$ and the measured observable $M$. 

Let us now consider the multiparameter case, discussed e.g., in ~\cite{gessnerNatComm20,du2025characterizingresourcesmultiparameterestimation}. Concretely, in our scenario this means that we want to  estimate $\vec r$ via the expectation values of a vector of observables $\vec M$.
The accuracy of such an estimator can be once again calculated via error-propagation and asymptotically can be bounded as (in a matrix form, at leading order in $1/\nu$):
\be
\mathrm{Cov}(\estr)
\gtrsim
\left(\nu \, \mathcal{M}_{\psi_0}(\vec H,\vec M) \right)^{-1} ,
\label{eq:moment_matrix_bound}
\ee
plus corrections that should vanish asymptotically in $1/\nu$.

Here, the right-hand side of \cref{eq:moment_matrix_bound} is given by the {\it moment matrix}
\be
\mathcal{M}_{\psi_0}(\vec H,\vec M):= C_{\psi_0}^T(\vec H, \vec M)\,\Gamma^S_{\psi_0}(\vec M)^{-1}\,C_{\psi_0}(\vec H, \vec M),
\ee
where $C_{\psi_0}$ is a Jacobian matrix that has the commutator form
\be
(C_{\psi_0})_{ij}
=
-i\,\av{[M_i,H_j]}_{\psi_0} .
\label{eq:jacobian_commutator}
\ee
Note that one should consider observables such that their commutators with the generators are nonvanishing in the probe state, thereby making the Jacobian matrix invertible, otherwise
it means that they are not sensitive to a change in (some of) the parameters.

Here, following on the illustration from the previous section, we focus on the probe states $\ket{\psi_0}$ that provide the overall minimal QCRB, i.e., they are solution of
the minimization problem in \cref{A_d}. As we mention, those states are such that the commutator between $Q$ and $P$ is nonvanishing, and this implies that the QCRB is not saturable.
At the same time, this also means that the canonical pair itself, besides being the vector of generators of the dynamics, is also a suitable vector of measurement observables, which are sensitive to both parameters.
Because of this, for those probe states we can also take the measured observables to coincide with the canonical pair itself, i.e., $\vec M = \vec R = \vec H$. 
In order to observe how such a scheme gets close to saturate the QCRB, we have to compare \cref{eq:moment_matrix_bound} with \cref{eq:qcrb_matrix}. In our concrete scenario we find
\be\label{eq:MoMvsQCRB}
\mathcal{M}_{\psi_0}(\vec R,\vec R) = \Omega^T_{\psi_0}(\vec R)\, \Gamma^S_{\psi_{0}}(\vec R)^{-1} \, \Omega_{\psi_0}(\vec R)  < 4\Gamma^S_{\psi_{0}}(\vec R) ,
\ee
where once again we have called $[\Omega_{\psi_0}(\vec R)]_{jk} = -i \av{[R_j,R_k]}_{\psi_0}= \mathrm{Im} [\Gamma_{\psi_0}(\vec R)]$ the commutator matrix of $\vec R$ onto the state $\ket{\psi_0}$.

Note that an alternative question could be to look for the probe state(s) that are 
such that {\it equality} is reached between the moment matrix and the QCRB, i.e.:
\be\label{eq:equalityMMQCRB}
\Omega^T_{\psi}(\vec M)\, \Gamma^S_{\psi}(\vec M)^{-1} \, \Omega_{\psi}(\vec M)  = 4 \Gamma^S_{\psi}(\vec R) ,
\ee
which however will necessarily involve other probe states and measurements, as we mentioned. In particular, one could look at the states that achieve the bound $\mathcal A_d^c$ in \cref{A_dc}, however necessarily with measurements different from $\vec R$.
This, together with a deeper analysis of the application of our framework to multiparameter estimation problems we leave for subsequent work.
It is worth mentioning here that in continuous variables, it is known that Gaussian states can saturate the condition in \cref{eq:equalityMMQCRB}~\cite{gessnerNatComm20,fadel2024quantummetrologycontinuousvariable,du2025characterizingresourcesmultiparameterestimation}.
 
Thus, we consider the states that are optimal in the unrestricted measurement scenario, namely those corresponding to values of $A_d$ in \cref{A_d}  and evaluate their performance under the restricted measurements $\vec R$ using \cref{eq:moment_matrix_bound}. First, we obtain a scalar form of the matrix inequality \cref{eq:moment_matrix_bound} as:
\be
\tr(\mathrm{Cov}(\estr))
\gtrsim \frac 1 \nu 
\mathcal A^M_d,
\label{eq:moment_matrix_bound_scalar}
\ee
where 
\be
\mathcal A^M_d=\tr\left(\left( \, \mathcal{M}_{\psi_0}(\vec R,\vec R) \right)^{-1}\right),
\label{eq:moment_matrix_bound_scalar2}
\ee
once again tracing the matrix inequality with the weight matrix $W=\id$. We then analyze how the resulting accuracy scales with the dimension. 
\begin{figure*}
    \includegraphics[width=0.7\linewidth]{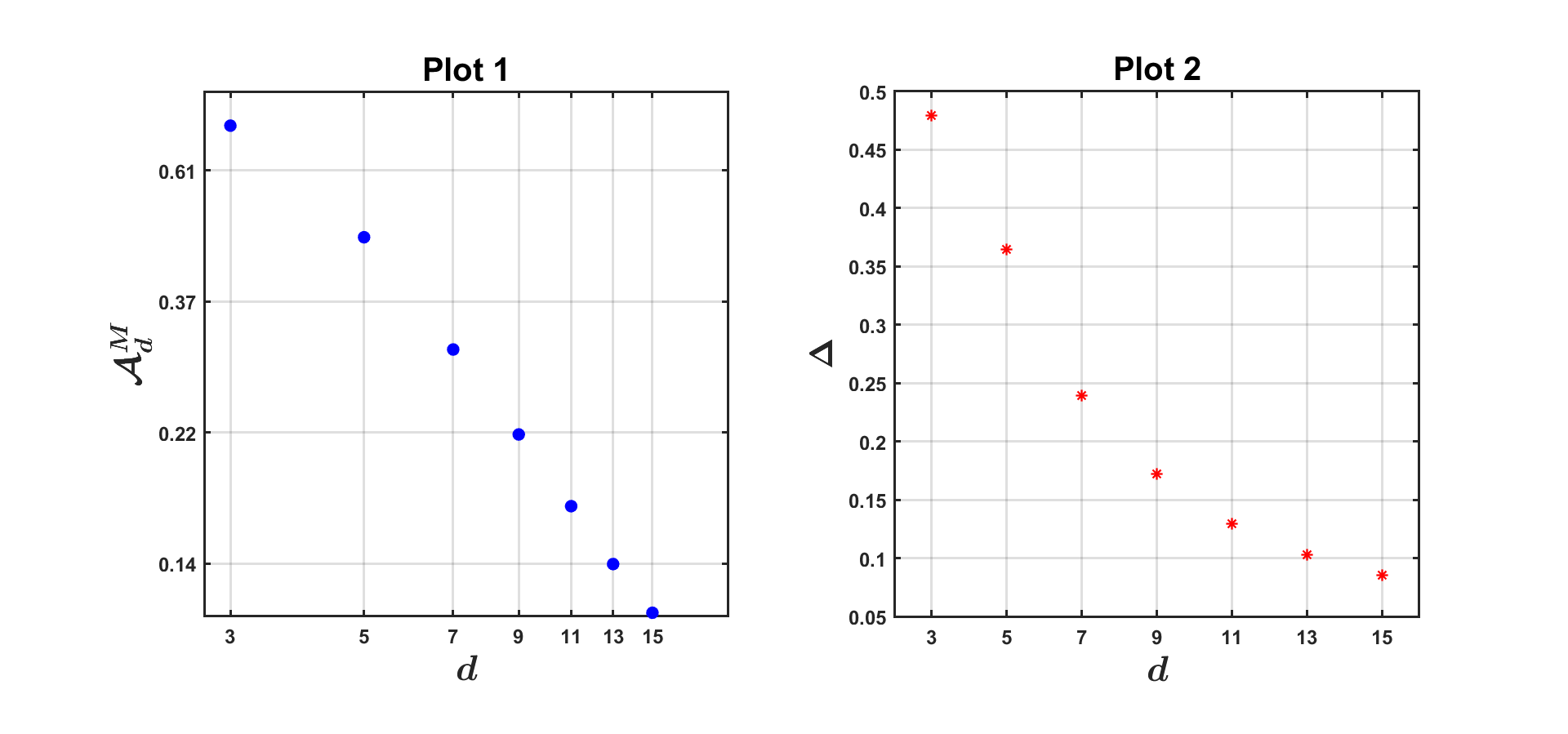}
    \caption{In plot $1$, the quantity $\mathcal A^M_d$ is shown as a function of the dimension on a log-log scale. This quantity provides a lower bound on the trace of the covariance matrix associated with the two estimators $\estr = (\estrcomp{1}, \estrcomp{2})$ (cf. \cref{eq:moment_matrix_bound_scalar}) with measurement $M=R$, and an initial state $\ket{\psi_0}$, chosen as the state that attains the minimal QCRB $\mathcal A_d$ (cf. \cref{A_d}). Plot 2 shows the difference, $\Delta=\mathcal A^M_d-\mathcal A_d$ as a function of the  dimension.}
    \label{fig:fig-accuracy vs dim 2 parameter with measurement}
\end{figure*}
Once again, as anticipated the accuracy increases with the dimension. In the  plot $2$ of Fig. \ref{fig:fig-accuracy vs dim 2 parameter with measurement}, we show the gap between the lower bounds to the accuracy corresponding to the restricted and unrestricted measurement scenarios i.e., $\Delta=\mathcal A^M_d-\mathcal A_d$ w.r.t. dimension. We observe that it shrinks with increasing dimension, indicating that the restricted setting with measurement $\vec{R}$ yields accuracies increasingly close to those of the optimal sensing scenario in higher dimensions.

Thus, following this argument, we observe how preparation uncertainties quantify the resource enabling sensitivity in a concrete multiparameter estimation scenario. 

\subsection{Covariance matrices and entanglement}
\label{sec:entanglement_criteria}

As a final application, let us consider a bipartite system with Hilbert space $\mathcal H_d \otimes \mathcal H_d$ and let us observe
how the analysis of the sigle-particle covariance matrix, ideally once again in the particular case of pure states, leads to a criterion for detecting entanglement.
Once again, for concreteness we focus on the particular case of (finite) $Q$ and $P$ variables, that we have discussed throughout the paper, 
but many of the statements will have more general validity.

Thus, let us consider the covariance matrix of the vector of single-particle observables $\vec R= (Q_1 , P_1 , Q_2 , P_2)$ where $\vec R_1 = (Q_1,P_1)$ and $\vec R_2 = (Q_2, P_2)$ are the local finite position and momentum operators. (Here, we use the simplified notation $Q_1 \equiv Q_1 \otimes \id$ and similarly for the other local observables, omitting to put the explicit tensor product with the identity.)
Such a covariance matrix takes the form
\be
\Gamma_\varrho(\vec R) = 
\left(\begin{array}{cc}
\Gamma(\vec R_1) & X(\vec R_1 , \vec R_2) \\
X^\dagger(\vec R_1 , \vec R_2) & \gamma(\vec R_2)
\end{array}\right),
\ee
where the blocks on the diagonal are given by the covariance matrices of the single particle canonical pairs and the off-diagonal blocks are given by
\be
    X(\vec R_1 , \vec R_2)  = \left(\begin{array}{cc}
\av{Q_1 \otimes Q_2} - \av{Q_1} \av{Q_2} & \av{Q_1 \otimes P_2} - \av{Q_1} \av{P_2} \\
\av{P_1 \otimes Q_2} - \av{P_1} \av{Q_2}  & \av{P_1 \otimes P_2} - \av{P_1} \av{P_2}
\end{array}\right) ,
\ee
which are essentially cross-covariances between the local observables.

It is well-known that the following matrix inequality holds for {\it all separable states}~\cite{WernerWolf2001,guhnecova,gittsovich08,LiuVitagliano2022}:
\begin{equation}
    \Gamma_\varrho(\vec R) \geq \kappa_1 \oplus \kappa_2 , \label{eq:DVCMCkappa}
\end{equation}
where a separable state is by definition any state that admits a decomposition of the form $\varrho_{\rm sep} = \sum_k p_k \ketbra{\psi^{(k)}_1} \otimes \ketbra{\psi^{(k)}_2}$, with $\ket{\psi^{(k)}_n}$ being pure single-particle states and $\{p_k\}$ being probabilities, and the bound is given by $\kappa_1 \oplus \kappa_2 \geq 0$ such that there exist ensembles of pure states $\{p_k,\ket{\psi_k^{(n)}}\}_k$ with $\kappa_n = \sum_k p_k \Gamma_{\psi^{(n)}_k}(\vec r_n)$.  
In other words, if a state is separable, there must exist such positive matrices $\kappa_1$ and $\kappa_2$, which are: (i) mixtures of covariance matrices of single particle pure states and also such that (ii) they give a lower bound to $\Gamma_\varrho(\vec R)$  as in \cref{eq:DVCMCkappa}. Otherwise, if no such matrices $\kappa_1$ and $\kappa_2$ exist, then the state $\varrho$ must be entangled.

From the matrix inequality in \cref{eq:DVCMCkappa} one can recover any scalar uncertainty-relation entanglement criterion~\cite{duan00entanglement,WernerWolf2001,hofman03,guhnecova,Hyllus06}. Once again, these can be obtained by tracing the matrix inequality with any positive semidefinite matrix $Z$ as $\tr( Z \Gamma_\varrho(\vec R)) \geq \tr( Z\kappa_1 \oplus \kappa_2 )$.
As a particularly relevant case we consider the weight matrix 
\begin{align}
Z = \begin{pmatrix}1&0&-1&0\\0&1&0&1\\-1&0&1&0\\0&1&0&1\end{pmatrix} ,
\end{align} 
and obtain the  criterion:
\be\label{eq:scalarentcrit_id}
\tr(Z\Gamma_\varrho(\vec R)) \geq \tr(Z(\kappa_1 \oplus \kappa_2) ) \geq 2 \min_\varrho \tr(\Gamma_\varrho(R_1)) ,
\ee
where the overall bound on the right-hand side can be obtained by looking for the minimal value of the trace of the single-particle covariance matrix, which can be found using our methods described earlier (cf~\cref{eq:U-groundstate-paper}). Explicitly, in this way we obtain a criterion analogous to the paradigmatic CV Simon-Duan uncertainty relation criterion~\cite{Simon2000,duan00}: 
\be\label{eq:Duan_like_cr}
\Var_{\varrho} (Q_1-Q_2)+\Var_\varrho(P_1+P_2) \geq 2 \min_{\psi \in \mathbb{C}^d} \tr(\Gamma_\psi(R_1)) := 2 U ,
\ee
where we called $U$ the bound on the right-hand side, which is the minimum possible trace of the single-particle covariance matrix of $(Q,P)$ for pure states.

\begin{figure}
    \begin{center}
\includegraphics[width=1.2\linewidth,trim=2.8cm 9cm 0cm 9cm,
clip]{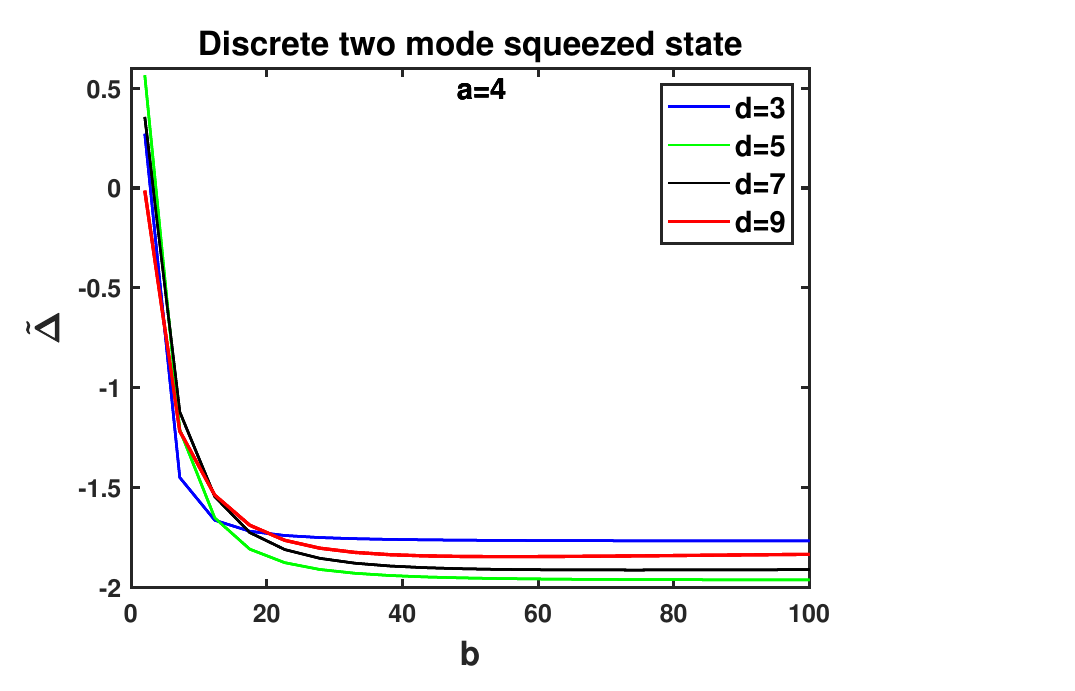}
    \caption{The difference $\tilde{\Delta}=\Var (Q_1-Q_2)+\Var(P_1+P_2) - 2U$ with respect to  the finite and discrete two mode squeezed states (\ref{eq:two_mode_disc_sq_state}) is plotted against real squeezing parameters $a=4$, $2\leq b\leq 100$ for different dimensions. The negative value of $\tilde{\Delta}$ implies violation of the separability criterion in \cref{eq:Duan_like_cr}, i.e., entanglement.}
    \label{fig:two_mode_sq_gauss}
    \end{center}
\end{figure}

Again, the inequality in \cref{eq:Duan_like_cr} must be valid for all separable states $\varrho$. The maximal violation of this criterion is achieved by the canonical maximally entangled state $\ket{\Phi}=\frac{1}{\sqrt{d}}\sum_{n_1,n_2}\ket{n_1,n_2}$, that is a simultaneous eigenstate of 
$Q_1-Q_2$ and $P_1+P_2$~\cite{hofman03}. For such a state, the left-hand side of \cref{eq:Duan_like_cr} vanishes in any dimension. This is analogous to the notion of the EPR state in continuous-variable systems~\cite{WANG2007}.
Another class of states that violates the above witness can be constructed in analogy with continuous-variable two-mode squeezed states:
\be\label{eq:two_mode_disc_sq_state}
\ket{\psi}\propto \sum_{n_1,n_2}\exp(-\frac{\pi}{d}\left(a(n_1-n_2)^2+\frac{(n_1+n_2)^2}{b}\right)) \ket{n_1,n_2},
\ee
where 
$d$ is the dimension and $a,b$ are analogous to squeezing parameters. In \cref{fig:two_mode_sq_gauss} we show how they violate \cref{eq:Duan_like_cr} for real values of $a,b$. In the limit $a,b\rightarrow \infty$ these states approach the maximally entangled state, just as two-mode squeezed states converge to the EPR state in the infinite squeezing limit \cite{WANG2007}.

Let us consider a noisy setting where the state is mixed with thermal noise. In particular we consider a thermal state of the Hamiltonian $H=(Q_1-Q_2)^2+(P_1+P_2)^2$, i.e., 
\be
\varrho_T = \frac{1} Z \exp\left( - \left( (Q_1-Q_2)^2 +(P_1+P_2)^2 \right)/T\right) , 
\ee
where $T$ is the temperature and $Z=\tr \exp\left( - \left( (Q_1-Q_2)^2 +(P_1+P_2)^2 \right)/T\right)$ is the partition function. At $T=0$, the state corresponds to a pure maximally entangled state. As $T$ increases, the state becomes increasingly more mixed, until eventually it becomes separable. At $T\rightarrow \infty$, it approaches the maximally mixed state. We evaluate the criterion in \cref{eq:Duan_like_cr} and identify the maximum temperature up $T^V_{thr}$ to which it can still detect entanglement. Comparing these   threshold temperatures,  with those obtained from existing separability criteria based on two mutually unbiased bases (MUB) \cite{Spengler2012ED,Morelli2023HDE}, denoted by $T^M_{thr}$, we find that our criterion consistently detects entanglement at higher temperatures. Specifically, for dimensions $d=3,5,7,$ $T^V_{thr}=2.05$,  whereas the MUB criterion detects entanglement only up to $T^M_{thr}=1.05$. For higher dimensions $d=9,11,13,15,$ the corresponding thresholds are $T^V_{thr}=1.05$ and $T^M_{thr}=0.05$. This is an interesting observation because MUB witnesses are known to be particularly effective for detecting maximally entangled states under added white noise \cite{Spengler2012ED}. However, this robustness does not seem to persist in the presence of thermal noise, where our bound exhibits greater resilience. 
\section{Conclusions and Outlook}
\label{sec:conclusions}

In conclusion, we have investigated the covariance-matrix uncertainty structure of a finite-dimensional canonical pair of observables, defined as two Fourier-conjugate analogues of position and momentum. Rather than focusing only on a single scalar uncertainty relation, we have characterized the set of attainable covariance matrices, in particular through its natural unitary invariants: trace and determinant. This provides a geometric description of the uncertainty region associated with finite-dimensional position and momentum, and makes explicit both the similarities with and the differences from the continuous-variable case.

As expected, finite-dimensional canonical pairs retain many of the structural features familiar from continuous-variable quantum mechanics, but at the same time, finite dimensionality introduces genuinely new features. The spectra of the observables are bounded, the set of attainable covariance matrices is compact, and uncertainty is constrained not only from below but also from above. 

Besides the foundational interest, we have analyzed how to connect the obtained uncertainty relations with operational tasks. In the metrological setting, the covariance matrix of the generators directly controls the quantum Fisher information for unitary parameter encodings, while additional restrictions on the accessible measurements lead to accuracy bounds expressed in terms of moment matrices and error-propagation relations~\cite{Helstrom1976,BraunsteinCaves1994,TothApellaniz2014,pezzerev18,gessnerPRL18}. In this sense, preparation uncertainty quantifies the resource enabling sensitivity, while measurement restrictions determine how much of this resource can be operationally accessed. Our results therefore suggest a systematic route for translating finite-dimensional uncertainty regions into achievable precision regions for multiparameter estimation, potentially extendable also to the continuous variable limit.

Concretely, a natural extension of the applications of our results in multiparameter estimation problems would include optimizing over probe states and measurements, studying the role of incompatibility in the encoding and in the measurement, investigating more deeply the best achievable accuracy, as compared to the ultimate bounds~\cite{Holevo1973}. A second direction is to include realistic noise, as well as analyzing the finite-dimensional variables resulting from CV systems after coarse graining, or finite-resolution detection. Since finite Fourier pairs can be interpreted as discretized or truncated analogues of continuous quadratures, this also opens the possibility of a quantitative comparison between discrete-variable, truncated continuous-variable, and genuine continuous-variable metrological protocols. Such a comparison could clarify, for example, which features of continuous-variable quantum enhancement survive finite resolution, and which are intrinsically tied to the infinite-dimensional limit.

The entanglement application provides another broad direction. We have shown how the single-particle uncertainty region for finite $Q$ and $P$ variables leads naturally to covariance-matrix separability criteria for bipartite finite-dimensional systems, including a discrete analogue of the Simon-Duan-type EPR criterion~\cite{duan00entanglement,hofman03}. This connects our construction with the general covariance-matrix criterion for separability and with local uncertainty relations~\cite{guhnecova,gittsovich08,GittsovichQuant10,LiuVitagliano2022}. In the examples considered here, the resulting witnesses detect maximally entangled states, including discrete analogues of two-mode squeezed states, in a way that is more robust to thermal noise than
other criteria based on mutually-unbiased bases, which can be seen as directly analogous using similar measurements. These examples indicate that finite-dimensional covariance witnesses can capture correlations that are naturally expressed in position--momentum language and may be competitive with other canonical witnesses.

A particularly interesting outlook is to reproduce, in finite dimension, the main structures of continuous-variable entanglement theory based on covariance matrices. In continuous-variable systems, covariance matrices play a central role in separability criteria, especially for Gaussian states, where they provide necessary and sufficient conditions~\cite{Simon2000,duan00entanglement,weedbrook2012gaussian,serafini2023quantum}. A finite-dimensional analogue of this theory would require identifying which parts of the Gaussian covariance-matrix formalism survive discretization, which require modification, and which fail because of finite-dimensional effects such as compactness and bounded spectra. This includes not only separability criteria, but also quantitative questions: covariance-matrix methods have been used to bound or quantify entanglement~\cite{Hyllus06,GiedkeetAlPRL2003,MarianandMarianPRA2008,MarianPRLEOF,GittsovichQuant10,FadelVitagliano_2021}, and it would be natural to investigate whether finite-dimensional position--momentum covariance data can be converted into computable lower bounds on entanglement monotones~\cite{vidal00,PlenioVirmani2007}. Such a program would be especially relevant for discrete approximations to Gaussian states and for truncated continuous-variable systems.

Beyond detecting strongly EPR-like entanglement, an important challenge is to improve the sensitivity of the method to broader classes of states. For example, it would be valuable to understand better how finite-dimensional covariance-matrix criteria perform with respect to other general methods, like those based on systematic construction of linear witnesses and positive but not completely positive maps, including the detection of positive-partial-transpose entangled states~\cite{gittsovich08,Hofmann_2003}. This may require going beyond the two-observable covariance matrix considered here, incorporating larger sets of local observables. The finite-dimensional setting is promising in this respect because the relevant optimization problems can often be formulated as finite semidefinite programs.

More generally, our results point toward a systematic comparison between discrete-variable and continuous-variable notions of uncertainty, metrological usefulness, and entanglement. Finite Fourier pairs provide a bridge between these regimes: they are finite-dimensional and experimentally accessible, but they retain a recognizable position--momentum structure. This makes them a useful platform for studying how continuous-variable concepts such as squeezing, EPR correlations, Gaussian covariance matrices, and phase-space geometry are modified by discretization and truncation. Extending the present analysis to multimode systems, many-body lattice models, and experimentally motivated coarse-grained quadrature measurements would further clarify the role of finite-dimensional uncertainty geometry in quantum information processing. 

\section*{Acknowledgments}
This research was funded in whole or in part by the Austrian Science Fund (FWF)  [\href{https://doi.org/10.55776/P35810}{10.55776/P35810}], [\href{https://doi.org/10.55776/P36633}{10.55776/P36633}] and the National Natural Science Foundation of China (No. 12405005). S.L. acknowledges the China Postdoctoral Science Foundation (No. 2023M740119). G.V. also acknowledges support from the Grant No. RYC2024-048278-I funded by MCIU/AEI/10.13039/501100011033 and FSE+.
 KS gratefully acknowledges funding from the projects DeQHOST APVV-22-0570, QUAS
VEGA 2/0164/25, Postdokgrant APD0161, and Stefan Schwarz programme.

\bibliography{Bibliography_UR.bib}

\appendix

\section{Minimum uncertainty states}\label{sec:minunc}

Here we quickly recap how to find states that saturate the uncertainty relation in  
\cref{eq:RobSchUR} in the case of pure states. \cref{eq:RobSchUR} can be seen as the Cauchy-Schwarz inequality for the two vectors
\be
\begin{aligned}
    \ket{\psi_A} &= (A- \av{A}_\psi ) \ket{\psi} := \shiftobs{A}{\psi} \ket{\psi}, \\
     \ket{\psi_B} &= (B- \av{B}_\psi )\ket{\psi} := \shiftobs{B}{\psi} \ket{\psi} .
\end{aligned}
\ee
In fact, we can see that it can be cast in the form
\be
\bra{\psi_A} \psi_A \rangle  \bra{\psi_B} \psi_B \rangle  \geq |\bra{\psi_A} \psi_B \rangle|^2 . 
\ee
This inequality is thus saturated by states that are such that the vectors $\ket{\psi_A}$ and $\ket{\psi_B}$ associated to the observables are parallel. 
This means that we have
\be\label{eq:satcondEq1}
\lambda \ket{\psi_A} = i \ket{\psi_B}  \implies  \shiftobs{A}{\psi} \ket{\psi} = i \shiftobs{B}{\psi} \ket{\psi} ,
\ee
and we then have
\be
\lambda = i \bra{\psi_A} \psi_B \rangle / \bra{\psi_A} \psi_A \rangle  = i \Cov{A,B}{\psi} / \var{A}{\psi} .
\ee
Furthermore, it can be easily seen that the states $\ket{\psi}$ solution to \cref{eq:satcondEq1} are such that~\cite{jackiwminuncJMP68,puriminimumuncPRA94}
\be\label{eq:satcondEq2}
\begin{aligned}
    \var{B}{\psi} - |\lambda|^2 \var{A}{\psi} &= - i \lambda \Re X_\psi , \\ 
   \var{B}{\psi} + |\lambda|^2 \var{A}{\psi} &= \lambda \Im X_\psi , 
\end{aligned}
\ee
where we called $X_\psi=\Cov{B,A}{\psi}$ for brevity.

Thus, as a result one obtains that those states can be labelled by two complex numbers $z$ and $\lambda$ and satisfy the equation:
\be\label{eq:sqcohstatesgen}
L(\lambda)\ket{z,\lambda}:=(\lambda A+ i B)\ket{z,\lambda} = z \ket{z,\lambda} ,
\ee
which is an eigenvalue equation for the operator $L(\lambda)$, where $\lambda$ is an arbitrary complex number and $z= \lambda\av{A} + i \av{B}$ is the corresponding (complex) eigenvalue.
Apart from these states, the only other states saturating \cref{eq:RobSchUR} are the eigenstates of $A$ (which have $\var{A}{}=0$ and thus also saturate the inequality). 

For these states, obtained for $\Re(\lambda) > 0$, the variances and covariance are determined by the commutator expectation as:
\be\label{eq:valuesofCovsforlambda}
\begin{aligned}
    \var{A}{z,\lambda}&=\frac{\av{C}_{z,\lambda}}{\Re \lambda} , \\
    \var{B}{z,\lambda}&=|\lambda|^2 \frac{\av{C}_{z,\lambda}}{\Re \lambda} , \\
    \Cov{A,B}{z,\lambda}&= - \av{C}_{z,\lambda} \frac{\Im \lambda}{\Re \lambda} ,
\end{aligned}
\ee
where
\be
C:=-\tfrac i 2 [A,B]
\ee
is the hermitean observable associated to the commutator. Thus, one can check that potentially for any value of $\lambda$, the uncertainty relation \eqref{eq:RobSchUR} is saturated~\footnote{Equivalently, one can reformulate the above statement by saying that states saturating the uncertainty relation in \cref{eq:RobSchUR} satisfy the equation
\be
(u L + v L^\dagger)\ket{z,u,v} = z \ket{z,u,v} ,
\ee
where
\be
L=A+iB ,
\ee
and $(u,v)$ is a pair of complex numbers. This latter formulation is actually a bit more general, as it also includes the eigenstates of $A$ and $B$.}.

Thus, the parameter $\lambda$ controls the degree of ``squeezing'' between the two observables' variances: $\lambda = 1$ corresponds to ``coherent'' states (i.e., balanced variances), while $\lambda \neq 1$ yields squeezed states with $\varho{B}/\varho{A} = \lambda^2$. 
The ratio $\varho{B}/\varho{A} = \lambda^2$ quantifies the generalized squeezing, while the commutator expectation sets the overall scale. The trace of the covariance matrix is
\be\label{eq:intelligent-trace}
\tr \Gamma_\psi = \av{-\tfrac{i}{2}[A,B]}_\psi \left(\lambda + \frac{1}{\lambda}\right).
\ee

\section{More details on the $d=3$ minimum uncertainty states}
\label{app:d3exampledetails}

Consider dimension $d=3$. We choose the computational basis $\{|-1\rangle, |0\rangle, |1\rangle\}$ and arrange column vectors in this order. The position and momentum operators are 
\bea
Q &= \sqrt{\frac{2\pi}{3}}\operatorname{diag}(-1,\,0,\,1) , \\
P &= \sqrt{\frac{2\pi}{3}}\begin{pmatrix} 0 & -\frac{i}{\sqrt{3}} & \frac{i}{\sqrt{3}} \\ \frac{i}{\sqrt{3}} & 0 & -\frac{i}{\sqrt{3}} \\ -\frac{i}{\sqrt{3}} & \frac{i}{\sqrt{3}} & 0 \end{pmatrix}.
\eea

From these, we define the annihilation operator
$$
A := Q + iP = \sqrt{\frac{2\pi}{3}}\begin{pmatrix} -\frac{1}{\sqrt{3}} & \frac{1}{\sqrt{3}} & -\frac{1}{\sqrt{3}} \\[4pt] -\frac{1}{\sqrt{3}} & 0 & \frac{1}{\sqrt{3}} \\[4pt] \frac{1}{\sqrt{3}} & -\frac{1}{\sqrt{3}} & \frac{1}{\sqrt{3}} \end{pmatrix}.
$$
This matrix is real but not symmetric.

Consider the unnormalized vector
$$
v_0 = \begin{pmatrix} 1 \\ 1+\sqrt{3} \\ 1 \end{pmatrix} ,
$$
the squared norm of which is
$|v_0|^2 = 6 + 2\sqrt{3}$. 

This corresponds to the discrete vacuum state, i.e., we have that
$$
|0_A\rangle = \frac{1}{\sqrt{6+2\sqrt{3}}}\,v_0 ,
$$
is an eigenstate of $A$ with zero eigenvalue, as it can be see via direct calculation.

Let us now define a finite-dimensional analogue of the squeezing operator. Firs, let us define the squeezing generator as
$$
K = \frac{1}{2}(QP + PQ).
$$
Explicit calculation yields
$$
K =2\pi \begin{pmatrix} 0 & \frac{i}{6\sqrt{3}} & 0 \\[4pt] -\frac{i}{6\sqrt{3}} & 0 & -\frac{i}{6\sqrt{3}} \\[4pt] 0 & \frac{i}{6\sqrt{3}} & 0 \end{pmatrix}.
$$
The matrix $L:= -iK$ is real and antisymmetric, so $e^{-i\xi K} = e^{\xi L}$ is a real orthogonal matrix.

We can also write $L = \theta J$ where $J$ is a normalized generator satisfying $J^3 = -J$ and $\theta$ represents the angular velocity. Since $L$ is antisymmetric, its eigenvalues are $\pm i\theta$ and $0$. Computing the characteristic polynomial gives
$$
\theta =2\pi \frac{\sqrt{6}}{18} = \frac{\sqrt{2}\pi}{3\sqrt{3}}.
$$
Setting $J = L/\theta$, we obtain
$$
J = \begin{pmatrix} 0 & \frac{\sqrt{2}}{2} & 0 \\[4pt] -\frac{\sqrt{2}}{2} & 0 & -\frac{\sqrt{2}}{2} \\[4pt] 0 & \frac{\sqrt{2}}{2} & 0 \end{pmatrix}.
$$
One can verify that $J^3 = -J$. Writing $\alpha = \theta \xi = \frac{\sqrt{2}\pi \xi}{3\sqrt{3}}$, the Rodrigues formula gives
$$
e^{\alpha J} = I + \sin\alpha\,J + (1 - \cos\alpha)\,J^2.
$$

Applying the squeezing operator to $v_0$, we compute
$$
v(\alpha) = e^{\alpha J}v_0 = v_0 + \sin\alpha\,(Jv_0) + (1-\cos\alpha)\,(J^2v_0), 
$$
which results in
$$
v(\alpha) = \begin{pmatrix} \cos\alpha + \frac{\sqrt{2}(1+\sqrt{3})}{2}\sin\alpha \\[4pt] (1+\sqrt{3})\cos\alpha - \sqrt{2}\sin\alpha \\[4pt] \cos\alpha + \frac{\sqrt{2}(1+\sqrt{3})}{2}\sin\alpha \end{pmatrix}.
$$
The normalized squeezed state family is then
$$
|\psi(\xi)\rangle = \frac{1}{\sqrt{6+2\sqrt{3}}} \begin{pmatrix} x(\alpha) \\ y(\alpha) \\ x(\alpha) \end{pmatrix},
$$
where
\bea
x(\alpha) &= \cos\alpha + \frac{\sqrt{2}(1+\sqrt{3})}{2}\sin\alpha, \\ 
y(\alpha) &= (1+\sqrt{3})\cos\alpha - \sqrt{2}\sin\alpha.
\eea

Let us now compute the covariance matrix along this family of states.
First, $|\psi(\xi)\rangle$ is a real vector in the chosen basis while $P$ is purely imaginary, so $\langle P\rangle = 0$. Second, the state vector has the form $(x, y, x)^T$, while the eigenvalues of $Q$ are symmetric about zero, so $\langle Q\rangle = 0$. 

The position variance is
$$
\operatorname{Var}(Q) = \langle Q^2\rangle = \frac{1}{|v_0|^2}\left(\frac{2\pi}{3}x^2 + 0 + \frac{2\pi}{3}x^2\right) = \frac{4\pi x^2}{3|v_0|^2}.
$$
Substituting the expression for $x(\alpha)$ and using the identities $\cos^2\alpha = \frac{1+\cos 2\alpha}{2}$, $\sin^2\alpha = \frac{1-\cos 2\alpha}{2}$, and $\sin\alpha\cos\alpha = \frac{\sin 2\alpha}{2}$, we expand
$$
x^2 = \cos^2\alpha + \sqrt{2}(1+\sqrt{3})\sin\alpha\cos\alpha + \frac{(1+\sqrt{3})^2}{2}\sin^2\alpha.
$$
Noting that $(1+\sqrt{3})^2 = 4 + 2\sqrt{3}$ and simplifying yields
$$
\operatorname{Var}(Q) =2\pi \left(\frac{1}{6} - \frac{\sqrt{3}}{18}\cos 2\alpha + \frac{\sqrt{6}}{18}\sin 2\alpha\right).
$$

For the momentum variance, after some straightforward algebra we obtain
$$
\operatorname{Var}(P) = \frac{4\pi(x-y)^2}{9|v_0|^2} ,
$$
and further computing $(x-y)^2$ gives
$$
\operatorname{Var}(P) = 2\pi\left(\frac{1}{6} - \frac{\sqrt{3}}{18}\cos 2\alpha - \frac{\sqrt{6}}{18}\sin 2\alpha\right).
$$

The trace of the covariance matrix is then given by
$$
\operatorname{Var}(Q) + \operatorname{Var}(P) = \frac{2\pi} {3}\left(1 - \frac{1}{\sqrt{3}}\cos 2\alpha\right) ,
$$
and substituting $\alpha = \frac{2\pi r}{3\sqrt{6}}$ we get
$$
\operatorname{Var}(Q) + \operatorname{Var}(P)  =\frac{2\pi}{3}\left( 1 - \frac{1}{\sqrt{3}}\cos\left(\frac{2\sqrt{2}\pi r}{3\sqrt{3}}\right)\right).
$$

Afterwards, for caclulating the determinant we need to calculate $\langle QP\rangle$ which is purely imaginary. A direct calculation results in
$$
\langle QP\rangle = \frac{4\pi ix(y-x)}{3\sqrt{3}|v_0|^2} ,
$$
where we get $x(y-x) = \sqrt{3}\cos^2\alpha - (3+2\sqrt{3})\sin^2\alpha$.
Substituting into the expression for $\langle QP\rangle$ we obtain with some simple algebra:
$$
\langle QP\rangle = \frac{2\pi i(3\cos 2\alpha - \sqrt{3})}{18} .
$$

Writing $\operatorname{Var}(Q)$ and $\operatorname{Var}(P)$ in terms of $\cos 2\alpha$ and $\sin 2\alpha$, their product takes the form of a difference of squares. Setting $a = \frac{1}{6} - \frac{\sqrt{3}}{18}\cos 2\alpha$ and $b = \frac{\sqrt{6}}{18}\sin 2\alpha$, we have $\operatorname{Var}(Q)\operatorname{Var}(P) = 4\pi^2 (a^2 - b^2)$. On the other hand, $|\langle QP\rangle|^2 = \frac{4\pi^2(3\cos 2\alpha - \sqrt{3})^2}{324}$. Direct expansion confirms that
$$
\operatorname{Var}(Q)\operatorname{Var}(P) = 4\pi^2\frac{(3\cos 2\alpha - \sqrt{3})^2}{324} = |\langle QP\rangle|^2 , 
$$
which implies $\tr \Gamma = 0$ for all values of $\xi$.

\section{Multiparameter quantum metrology and the Holevo Cram\'er-Rao bound}\label{app:multi_metrology}

In this appendix we briefly summarize the basic notions of multiparameter quantum estimation theory, specialized to the case of a unitary encoding. We consider a smooth family of states
$\varrho_{\vec\theta}$ depending on a real parameter vector $\vec\theta=(\theta_1,\dots,\theta_M)^T\in\mathbb{R}^M$,
encoded by a unitary channel $\varrho_{\vec\theta}=U(\vec\theta)\varrho_0 U^\dagger(\vec\theta)$. 
A POVM is repeated $\nu$ times independently, and from the outcomes one constructs an estimator vector $\vec\Theta$ for $\vec\theta$.

A POVM $\{\Pi_x\}$ with outcome distribution
$p(x|\vec\theta)=\tr[\varrho_{\vec\theta}\Pi_x]$
induces the (classical) Fisher information matrix (FIM)
\begin{equation}
\big[I(\vec\theta)\big]_{\mu\nu}
=
\sum_x p(x|\vec\theta)\,
\partial_\mu \ln p(x|\vec\theta)\,
\partial_\nu \ln p(x|\vec\theta),
\end{equation}
where for simplicity we use the notation $\partial_\mu := \frac{\partial}{\partial\theta_\mu}$.
For any (locally) unbiased estimator $\vec\Theta$,
$\partial_\mu \mathbb{E}_{\vec\theta}[\Theta_\nu]=\delta_{\mu\nu}$,
the classical Cram\'er--Rao bound gives the matrix inequality
\begin{equation}
{\rm Cov}(\vec\Theta)\ \succeq\ \frac{1}{\nu}\, I^{-1}(\vec\theta).
\label{eq:classical_crb_app}
\end{equation}
In practice, one often evaluates \eqref{eq:classical_crb_app} through a positive semidefinite weight matrix $W\succeq 0$,
\begin{equation}
\tr\big(W\,{\rm Cov}(\vec\Theta)\big)\ \ge\ \frac{1}{\nu}\,\tr\big(W\, I^{-1}(\vec\theta)\big),
\label{eq:weighted_classical_crb_app}
\end{equation}
which corresponds to a scalar figure of merit that fixes how different parameters are traded off.

A further bound that takes into account the classical FIM arising over arbitrary measurements yields the {\it quantum Fisher information matrix} (QFIM). This is 
defined in terms of the {\it symmetric logarithmic derivatives} SLDs $\{L_\mu\}_{\mu=1}^M$, that are such that
\begin{equation}
\partial_\mu \varrho_{\vec\theta}
=
\frac12\Big(L_\mu\varrho_{\vec\theta}+\varrho_{\vec\theta}L_\mu\Big),
\qquad
L_\mu = L^{\dagger}_\mu,
\label{eq:sld_def_app}
\end{equation}
and the corresponding QFIM is
\begin{equation}
\big[F_{\varrho_{\vec\theta}}^Q\big]_{\mu\nu}
:=
\frac12\,\tr\Big(\varrho_{\vec\theta}\{L_\mu,L_\nu\}\Big)
=
\Re{\tr\Big(\varrho_{\vec\theta} L_\mu L_\nu\Big)}.
\label{eq:sld_qfim_app}
\end{equation}
The associated (matrix) quantum Cram\'er--Rao bound (QCRB) reads
\begin{equation}
{\rm Cov}(\vec\Theta)\ \succeq\ \frac{1}{\nu}\,\big(F_{\varrho_{\vec\theta}}^Q\big)^{-1}.
\label{eq:sld_qcrb_app}
\end{equation}
One can convert \eqref{eq:sld_qcrb_app} into a scalar inequality by tracing with $W\succeq 0$:
\begin{equation}
\tr\big(W\,{\rm Cov}(\vec\Theta)\big)\ \ge\ \frac{1}{\nu} \tr\Big(W\,\big(F_{\varrho_{\vec\theta}}^Q\big)^{-1}\Big).
\label{eq:sld_weighted_app}
\end{equation}

In the single-parameter case, the QCRB is (asymptotically) saturable by an appropriate measurement. In the multiparameter case this need not hold, because a measurement optimal for one component can be incompatible with one optimal for another. A standard asymptotic \emph{compatibility} condition for simultaneous attainability of \eqref{eq:sld_qcrb_app} is the vanishing of the \emph{average} SLD commutators,
\begin{equation}
\left[J_{\varrho_{\vec\theta}}\right]_{\mu\nu}
:=
\frac{1}{2i}\tr\Big(\varrho_{\vec\theta}[L_\mu,L_\nu]\Big)
=
\Im{\tr\Big(\varrho_{\vec\theta}L_\mu L_\nu\Big)}
=0
\quad \forall\,\mu,\nu .
\label{eq:sld_compatibility_app}
\end{equation}
When \eqref{eq:sld_compatibility_app} fails, the SLD-QCRB remains a valid lower bound but is generally not tight.

For unitary encodings $\varrho_{\vec\theta}=U(\vec\theta)\varrho_0U^\dagger(\vec\theta)$ these conditions can be expressed in terms of the (Hermitian) local generators
\begin{equation}
H_\mu(\vec\theta)
:=
i\,U^\dagger(\vec\theta)\,\partial_\mu U(\vec\theta) .
\label{eq:generator_def_app}
\end{equation}
For pure probes $\varrho_{\vec\theta}=\ket{\psi_{\vec\theta}}\!\bra{\psi_{\vec\theta}}$, the QFIM is given by
\begin{equation}
F_{\varrho_{\vec\theta}}^Q = 4\Gamma_{\varrho_{\vec\theta}}^S(\vec H) ,
\label{eq:pure_qfim_app}
\end{equation}
where in the right-hand side there is the symmetric (or real part of) covariance matrix of the generators.
The saturability condition of the matrix QCRB becomes
\begin{equation}
\big[J_{\varrho_{\vec\theta}}\big]_{\mu\nu}
= -2i\tr\Big(\varrho_{\vec\theta}[H_\mu,H_\nu]\Big)
=
-4\Im{\tr\Big(\varrho_{\vec\theta}H_\mu H_\nu\Big)} =0 ,
\label{eq:pure_qfim_app}
\end{equation}
which is then given in terms of the imaginary part of the covariance matrix of the generators~\cite{liu2020quantum}.

Note that for pure states the QFIM and the maatrix $J$ are proportional to the metric tensor and the Berry curvature matrix~\cite{liu2020quantum}.

\section{Details on the method-of-moments estimator}
\label{app:MoMdetails}

In this appendix we recall the method-of-moments estimator in the form used in the main text.  
First let us recall the single-parameter case.
Let $\{E_q\}$ be a PVM or, more generally, a POVM with outcomes $q$. The outcome probability is given by the usual Born rule:
\begin{equation}
{\rm p}(q|\theta)=\tr[\varrho_\theta E_q] .
\end{equation}
We associate to each outcome a real number $m(q)$ and define the theoretical mean
\begin{equation}
\mu(\theta):=\mathbb{E}_\theta[M]
=\sum_q m(q)\,{\rm p}(q|\theta).
\end{equation}
For $\nu$ independent repetitions, the empirical mean is
\begin{equation}
\bar M_\nu:=\frac{1}{\nu}\sum_{t=1}^{\nu}m(q_t).
\end{equation}
The method of moment estimator is defined by matching the empirical and theoretical moments,
\begin{equation}
\mu(\Theta)=\bar M_\nu .
\end{equation}
Assuming that $\mu(\theta)$ is locally invertible, a first-order expansion around the true value gives
\begin{equation}
\Theta-\theta
\simeq
\frac{\bar M_\nu-\mu(\theta)}{\partial_\theta\mu(\theta)} .
\end{equation}
By the central limit theorem,
\begin{equation}
\sqrt{\nu}\bigl(\bar M_\nu-\mu(\theta)\bigr)
\longrightarrow
\mathcal{N}\bigl(0,\Var_\theta M\bigr),
\end{equation}
and hence the estimator is asymptotically unbiased, with leading mean-squared error
\begin{equation}
\mathrm{MSE}[\Theta]
=
\frac{1}{\nu}
\frac{\Var_\theta M}{\bigl(\partial_\theta\mu(\theta)\bigr)^2}
+O(\nu^{-2}).
\label{eq:mom_single_mse_app}
\end{equation}
If the statistic is associated with measuring a Hermitian observable $M$ on $\varrho_\theta$, then
\begin{equation}
\mu(\theta)=\langle M\rangle_{\varrho_\theta},
\qquad
\Var_\theta M=\Var_{\varrho_\theta} M ,
\end{equation}
are calculated from quantum theory.

Let us now consider the multiparameter case~\cite{pezzerev18,gessnerPRL18,du2025characterizingresourcesmultiparameterestimation}.
Let $\vec\theta=(\theta_1,\dots,\theta_r)^T$ be the vector of unknown parameters, and let a single measurement produce an outcome $q$ from which we extract a vector of real statistics
\begin{equation}
\vec M(q)=\bigl(m_1(q),\dots,m_k(q)\bigr)^T .
\end{equation}
The empirical and theoretical moment vectors are
\begin{equation}
\bar{\vec M}_\nu
=
\frac{1}{\nu}\sum_{t=1}^{\nu}\vec M(q_t),
\qquad
\vec\mu(\vec\theta)
=
\mathbb{E}_{\vec\theta}[\vec M].
\end{equation}
The method of moments estimator is defined by
\begin{equation}
\vec\mu(\vec\Theta)=\bar{\vec M}_\nu .
\label{eq:mom_multi_def_app}
\end{equation}
Introducing the Jacobian
\begin{equation}
C_{ij}(\vec\theta)
:=
\partial_{\theta_j}\mu_i(\vec\theta),
\label{eq:C_def_app}
\end{equation}
the multivariate central limit theorem gives
\begin{equation}
\sqrt{\nu}\bigl(\bar{\vec M}_\nu-\vec\mu(\vec\theta)\bigr)
\longrightarrow
\mathcal{N}\bigl(0,\Gamma_{\vec\theta}(\vec M)\bigr),
\end{equation}
where $\Gamma_{\vec\theta}(\vec M)$ is the covariance matrix of the sampled probability.  Linearizing \eqref{eq:mom_multi_def_app} yields the asymptotic error matrix
\begin{equation}
\mathrm{MSE}[\vec\Theta]
=
\frac{1}{\nu}
\Bigl(
C^T\Gamma_{\vec\theta}^{-1}(\vec M)\,C
\Bigr)^{-1}
+O(\nu^{-2}),
\label{eq:mom_multi_mse_app}
\end{equation}
whenever the inverse exists.  Equivalently, the leading sensitivity is determined by the moment matrix as $\mathrm{MSE}[\vec\Theta]
=
\frac{1}{\nu}\,
\mathcal{M}_{\vec\theta}^{-1}(\vec M)
+O(\nu^{-2})$, with
\begin{equation}\label{eq:moment_matrix_mom_app}
\mathcal{M}_{\vec\theta}(\vec M)
:=
C^T\Gamma_{\vec\theta}^{-1}(\vec M)\,C .
\end{equation}

For a unitary parameter encoding
\begin{equation}
\varrho_{\vec\theta}
=
U(\vec\theta)\varrho\,U^\dagger(\vec\theta),
\end{equation}
we may work in the Heisenberg picture and write
\begin{equation}
M_i(\vec\theta)
=
U^\dagger(\vec\theta)M_iU(\vec\theta) .
\end{equation}
Then
\begin{equation}
C_{ij}(\vec\theta)
=
\partial_{\theta_j}\langle M_i(\vec\theta)\rangle_{\varrho}
=
-{\rm i}
\bigl\langle
[M_i(\vec\theta),H_j(\vec\theta)]
\bigr\rangle_{\varrho} ,
\label{eq:C_commutator_app}
\end{equation}
where we introduced the generators $H_j(\vec\theta) = {\rm i}\,U^\dagger(\vec\theta)\partial_{\theta_j}U(\vec\theta)$ as in \cref{eq:generator_def_app}.

Thus, for jointly measurable observables, the matrix $C$ is the commutator matrix appearing in the main text, while $\Gamma_{\vec\theta}(\vec M)$ is the covariance matrix estimated directly from the same experimental data.  This gives the moment-matrix expression used in the main text,
\begin{equation}
\mathcal{M}_{\varrho_{\vec\theta}}(\vec M)
=
\Omega^T_{\varrho_{\vec\theta}}(\vec M,\vec H)\,
\Gamma^{-1}_{\varrho_{\vec\theta}}(\vec M)\,
\Omega_{\varrho_{\vec\theta}}(\vec M,\vec H),
\label{eq:moment_matrix_commutator_app}
\end{equation}
with the corresponding asymptotic relation
\begin{equation}
\mathrm{MSE}[\vec\Theta]
=
\frac{1}{\nu}
\mathcal{M}^{-1}_{\varrho_{\vec\theta}}(\vec M)
+O(\nu^{-2}).
\label{eq:MSEfromMOM_app}
\end{equation}

When the observables $M_i$ commute and are measured through a common PVM, each experimental run produces a joint value of all components of $\vec M$.  In that case, the empirical covariance matrix converges to the quantum covariance matrix
\begin{equation}
\bigl[\Gamma_{\varrho_{\vec\theta}}(\vec M)\bigr]_{ij}
=
\langle M_iM_j\rangle_{\varrho_{\vec\theta}}
-
\langle M_i\rangle_{\varrho_{\vec\theta}}
\langle M_j\rangle_{\varrho_{\vec\theta}} .
\label{eq:Gamma_commuting_app}
\end{equation}
For non-commuting observables, the same formal expression is replaced by the real, symmetrized covariance
\begin{equation}
\bigl[\Gamma^S_{\varrho_{\vec\theta}}(\vec M)\bigr]_{ij}
=
\frac{1}{2}
\langle M_iM_j+M_jM_i\rangle_{\varrho_{\vec\theta}}
-
\langle M_i\rangle_{\varrho_{\vec\theta}}
\langle M_j\rangle_{\varrho_{\vec\theta}} .
\label{eq:Gamma_sym_app}
\end{equation}
However, non-commuting observables cannot in general be associated with a single sharp joint measurement.  A joint measurement must then be described by a generally non-projective POVM, and the covariance entering \eqref{eq:moment_matrix_mom_app} is the covariance of the resulting classical data stream, which may contain additional measurement noise~\cite{Busch14}.

\end{document}